\documentclass{aa}  
\usepackage{graphicx}
\usepackage{txfonts}
\usepackage{bm}
\usepackage[authoryear]{natbib}
\bibpunct{(}{)}{;}{a}{}{,}
\begin{document}
   \title{A fast method for Stokes profile synthesis}

   \subtitle{Radiative transfer modeling for ZDI and Stokes profile inversion}
   
   \author{T. A. Carroll,
          M. Kopf,
	  \and
	  K. G. Strassmeier 
          }

   \offprints{T. A. Carroll}

   \institute{Astrophysikalisches Institut Potsdam,
              An der Sternwarte 16, D-14482 Potsdam, Germany
	      \email{tcarroll@aip.de}
             }

   \date{Received May 2, 2006; accepted June 5, 2007}

% \abstract{}{}{}{}{} 
% 5 {} token are mandatory
 
  \abstract
   {The major challenges for a fully polarized radiative transfer driven approach 
   to Zeeman-Doppler imaging are still the enormous computational requirements.
   In every cycle of the iterative interplay between the 
   forward process (spectral synthesis) and the inverse process (derivative based optimization) 
   the Stokes profile synthesis requires several thousand evaluations of the 
   polarized radiative transfer equation for a given stellar surface model.}
   {To cope with these computational demands and to allow for
   the incorporation of a full Stokes profile synthesis
   into Doppler- and Zeeman-Doppler imaging applications as well as into 
   large scale solar Stokes profile inversions, 
   we present a novel fast and accurate 
   synthesis method for calculating local Stokes profiles.}
   {Our approach is based on artificial neural network models, which we use to 
   approximate the complex non-linear mapping between the most important 
   atmospheric parameters and the corresponding Stokes profiles. 
   A number of specialized artificial neural networks, are 
   used to model the functional relation between the model atmosphere, 
   magnetic field strength, field inclination, and field azimuth, on one hand
   and the individual components $(I,Q,U,V)$ of the Stokes profiles, on the other hand.}
   {We performed an extensive statistical evaluation and show that
   our new approach yields accurate local as well as disk-integrated Stokes profiles 
   over a wide range of atmospheric conditions. The mean rms errors for the Stokes $I$
   and $V$ profiles are well below 0.2 \% compared to the exact numerical solution.
   Errors for Stokes $Q$ and $U$ are in the range of 1 \%. Our approach does not only offer
   an accurate approximation to the LTE polarized radiative transfer
   it, moreover, accelerates the synthesis by a factor of more than 1000.}
  {}


   \maketitle
%
%________________________________________________________________

\section{Introduction}
   The successful application of Doppler- and Zeeman-Doppler
   imaging (DI \& ZDI) in the last two decades has enormously contributed to
   our current knowledge about stellar surface activity and magnetism
   \citep{Strass98, Donati99b, Pisk02, Strass02, Donati03, Koch04, Donati06}.
   Though it has its limitation, Zeeman-Doppler imaging is the only available method 
   that allows to estimate the surface distribution of stellar magnetic fields. 
   Doppler- and Zeeman-Doppler imaging rely on the complex and computational expensive 
   interplay between the forward process of spectral line synthesis on the one hand and the 
   inverse process of a regularized optimization or fitting procedure on the other hand.
   Based on a highly discretized surface model Doppler- and Zeeman-Doppler imaging
   iteratively seeks the best possible solution of the surface temperature and/or
   magnetic field surface distribution, which is compatible with the observed
   rotationally-modulated line profiles.
   
   The spectral synthesis plays a decisive role here and requires a
   good knowledge of accurate atomic data, continuous as well as line opacities,
   broadening mechanisms, model atmospheres etc. 
   The usual way to cope with the synthesis in DI applications, which allows for the calculation 
   of disk-integrated line profiles in a reasonable time frame, is the 
   compilation of a precalculated database of local line profiles 
   for large number of different atmospheric models and lines-of-sight (LOS) \citep{Rice00}. 
   But unlike DI, the radiative 
   transfer in magnetized media requires the solution of the coupled 
   set of equation for all four Stokes parameters \citep{Rees87} and moreover, since the magnetic field
   is a vector quantity, a proper precalculation quickly becomes prohibitive for ZDI applications in
   the general case where no simple global magnetic field organization is present. 
   
   Due to these complexities and enormous demands many applications of ZDI (but also DI) 
   still rely on approximate methods or even completely avoid a radiative transfer modeling approach.
   Some use an ad-hoc line synthesis where Stokes $I$ profiles are used from template
   stars or modeled by simple Gaussian profiles \citep{Donati97a}, from which the corresponding 
   Stokes $V$ profiles are derived by means of the weak-field approximation \citep{Donati97b,Donati99a}.
   In this respect, one is able to entirely bypass radiative transfer modeling.   
   
   In an effort to provide the necessary tools for a fully polarized radiative transfer
   driven approach to ZDI as well as to provide the diagnostic capabilities for the next
   generation of a high-resolution spectropolarimeter (PEPSI) at the 8.4 m Large Binocular Telescope (LBT) 
   \citep{Strass03,Strass08}, we developed the Zeeman-Doppler imaging code \emph{iMap} \citep{Carroll07}.
   To accelerate the synthesis process, our ZDI code incorporates, besides the conventional 
   numerical implementation of the polarized radiative transfer, a novel
   fast and accurate method for calculating local and disk-integrated synthetic Stokes profiles under
   local thermodynamic equilibrium (LTE). 
   This approach uses artificial neural networks (ANNs) to approximate the polarized radiative transfer 
   calculations.
   Artificial neural networks have already proven to be a fast alternative for solar Stokes profile inversions and 
   Zeeman tomography \citep{Carroll01, Socas05a, Socas05b, Lites07, Carroll08}. In the here presented approach
   ANNs are trained on the basis
   of the numerical solution of the polarized radiative transfer equation to find an
   approximate representation of the underlying complex mapping between atmospheric quantities like temperature, 
   magnetic field, velocity, etc., and the corresponding Stokes profiles.
   A detailed assessment and evaluation of the ANN synthesis
   confirms the accuracy of our novel approach 
   and moreover demonstrates that the ANN approach can
   accelerate the synthesis process by three orders of magnitude.
   
   This paper is organized as follows: Since our method strongly depends on the 
   accuracy of the underlying numerical method used for the calculation of synthetic Stokes spectra, we
   first perform an in-depth analysis and assessment of our polarized radiative transfer code in Sect. \ref{Sect:2}. 
   For this purpose, we made a detailed comparison between our code \emph{iMap} and the C{\small OSSAM} synthesis code 
   \citep{Stift00}, which was 
   also part of the benchmark test and inter-agreement analysis of \citet{Wade01}. In Sect.
   \ref{Sect:3}, we give a short introduction into the ANN models we used in this work and
   describe the calculation of the training database for different
   atmospheric models, magnetic field configurations and Zeeman sensitive spectral lines. 
   In Section \ref{Sect:4} we evaluate the accuracy and performance of the trained ANNs relative to the 
   exact numerical solution.
   Sect. \ref{Sect:5} presents a benchmark test that demonstrates the exceptional speed of the new synthesis method 
   compared to the conventional numerical procedure.
   Finally, we conclude in Sect. \ref{Sect:6} with a discussion and summary of the
   presented fast Stokes profile synthesis.

%__________________________________________________________________

\section{Numerical polarized radiative transfer}
\label{Sect:2}

Because of the particular importance of an accurate Stokes profile synthesis for this work, 
we first give a brief overview of the basics of the polarized line formation under
local thermal equilibrium (LTE)
as well as a short introduction of the polarized radiative transfer module implemented within our ZDI code \emph{iMap}.
This is followed by a thorough benchmark test with the existing polarized radiative transfer code C{\small OSSAM}
\citep{Stift00}.

\subsection{The transport equation}

The basic transport equation for the Stokes vector $\vec{I}$ in the frequency $\nu$ reads,
\begin{equation}
  \mu \frac{d\vec{I}_{\nu}}{dz} = -\vec{K}_{\nu}(\vec{I}_{\nu} - \vec{S}_{\nu}) \; ,
  \label{Eq:1}
\end{equation}
where $\mu \equiv$ cos ($\theta$) and theta the angle between the surface normal and the LOS.
$\vec{S}_{\nu}$ is the source vector and $\vec{K_{\nu}}$ the propagation matrix, which contains
contributions from true absorption and anomalous dispersion. For the sake of a better
readability, we omit the index for the frequency in the following.
Assuming local thermodynamic equilibrium (LTE) and neglecting continuum polarization
the source vector is given by
\begin{equation}
  \vec{S} = B(T) \: \vec{e}_0    \; ,
  \label{Eq:2}
\end{equation}
where $B(T)$ is the Planck function at local temperature $T$ and
$\vec{e_0} = (1,0,0,0)^{T}$ .
The total absorption or propagation matrix is given by
\begin{equation}
   \vec{K} = \kappa_c \vec{1} + \kappa_0 \vec{\Phi} \; .
   \label{Eq:3}
\end{equation}
Here, $\kappa_c$ is the continuum absorption coefficient, $\kappa_0$ the 
line absorption coefficient, $\vec{1}$ denotes the 4 $\times$ 4 identity matrix, and
$\vec{\Phi}$ the line propagation matrix, given as
\begin{eqnarray}
   \vec{\Phi} & = & \left(
        \begin{array}{cccc}
           \phi_I  &  \phi_Q   &  \phi_U  &  \phi_V \\
           \phi_Q  &  \phi_I   &  \phi'_V &  -\phi'_U \\
           \phi_U  & -\phi'_V  &  \phi_I  &  \phi'_Q \\
           \phi_V  &  \phi'_U  & -\phi'_Q &  \phi_I 
        \end{array} 
        \right) \; ,
  \label{Eq:4}
\end{eqnarray}
where
\begin{equation}
  \begin{array}{ccl}
    \phi_I & = & \frac{1}{2} \phi_p \: \sin^2\gamma + \frac{1}{4} (\phi_r + \phi_b) (1 + \cos^2\gamma) \\
    \phi_Q & = & \frac{1}{2} [\phi_p - \frac{1}{2} (\phi_r + \phi_b)]\: \sin^2 \gamma \: \cos 2\chi \\
    \phi_U & = & \frac{1}{2} [\phi_p - \frac{1}{2} (\phi_r + \phi_b)]\: \sin^2 \gamma \: \sin 2\chi \\
    \phi_V & =  &\frac{1}{2} (\phi_r - \phi_b) \cos\gamma \\
    \phi'_Q & = & \frac{1}{2} [\phi'_p - \frac{1}{2} (\phi'_r + \phi'_b)]\: \sin^2\gamma \: \cos 2\chi \\
    \phi'_U & = & \frac{1}{2} [\phi'_p - \frac{1}{2} (\phi'_r + \phi'_b)]\: \sin^2\gamma \: \sin 2\chi \\
    \phi'_V & = & \frac{1}{2} (\phi'_r - \phi'_b) \cos\gamma \; .
  \end{array}
  \label{Eq:5}
\end{equation}
The inclination angle of the magnetic field vector relative to the 
LOS is denoted herein as $\gamma$ and the azimuthal angle of the field vector as $\chi$ . 
The absorption profiles $\phi_{p,b,r}$ and the anomalous dispersion profiles
$\phi'_{p,b,r}$ are given by the Voigt- and Faraday-Voigt function respectively 
\citep[see, e.g.,][]{Rees87} and are weighted by the normalized strength of the respective atomic transitions 
\citep[see, e.g.,][]{Stenflo94}.

\subsection{Physical foundations}

Our LTE code (the forward module of \emph{iMap}) works on the basis of a given model atmosphere, e.g., provided by
A{\small TLAS}9 \citep{Kurucz93} or Phoenix \citep{Haus99} or by semi-empirical models \citep{Holweger74}.
The program reads atomic line parameters, which are extracted from the Vald database \citep{Pisk95,Kupka99}.
The calculations of the continuous opacities rest on a code written by \citet{Witt74} which accounts 
for free-free and bound-free transitions of H$^{-}$, \ion{H}{i}, \ion{He}{i}, He$^{-}$, H$^{-}_2$,
H$^{+}_2$ as well as for the metals C, Na, Mg. Scattering coefficients are computed for Rayleigh scattering
by neutral hydrogen, molecular hydrogen, and neutral helium, and for Thomson scattering by free electrons. 
Radiation damping and Stark broadening are taken into account as well as collisional broadening (van der Waals) 
according to \citet{Anstee95}.
Partition functions are calculated according to \citet{Bolton70} and \citet{Aller72}. Abundances, ionization energies, and
partition functions are supplied for 83 elements \citep{Witt74}. 
The Zeeman patterns are calculated under $\vec{L}$-$\vec{S}$ coupling. Voigt- and Faraday-Voigt functions are evaluated
using the rational approximation method of \citet{Humli82}.

The numerical integration of the polarized radiative transfer is provided by a
diagonal element lambda operator (DELO) method \citep{Rees89} and accounts for magneto-optical effects.
The method is implemented in its linear (original) form as well as in the quadratic version, which 
uses a parabolic approximation of the source function \citep{Socas00}.

For the disk-integration and calculation of stellar (flux) spectra, the underlying surface model in 
\emph{iMap} is parameterized on a variable
equal-area or equal-degree partition with a minimum element (pixel) size of 1 $\times$ 1 degree. For each surface
element, a local Stokes vector is calculated with respect to its position and underlying atmospheric parameters
(abundance, Doppler velocity, bulk velocity, temperature and pressure structures,
magnetic field, micro- and macroturbulence). All atmospheric
parameters are also allowed to vary along their local vertical
direction, which facilitates the complete description of a
3-dimensional atmosphere. Limb darkening is fully accounted for by
adjusting the depth stratification of the model atmosphere (temperature and pressure) 
for each surface element with respect to the LOS, i.e., recalculation of the optical depth scale
according to $\mu$ (angle between the local normal vector and the LOS).
Field structures on the surface (in temperature, abundance and magnetic fields) can either be described by
setting individual surface elements or using spherical harmonics.
\begin{figure*}[t*]
\centering
\includegraphics[width=8cm,height=7cm]{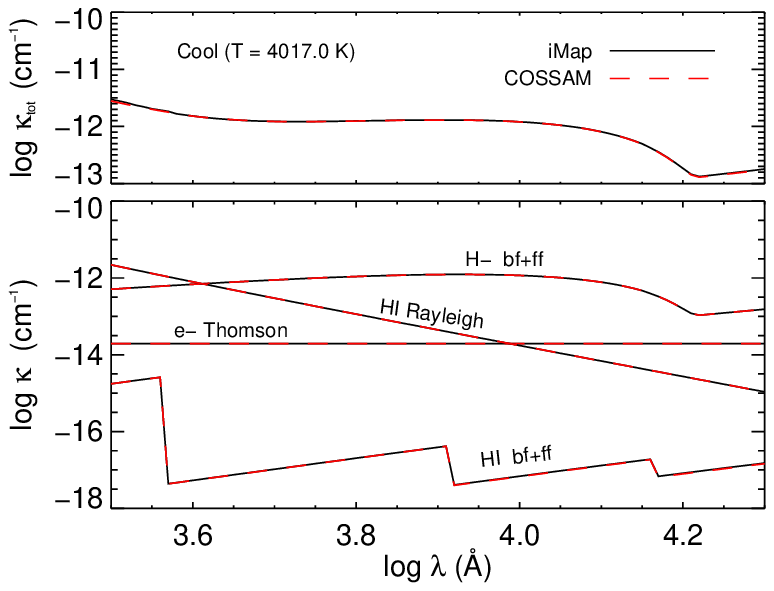}
\includegraphics[width=8cm,height=7cm]{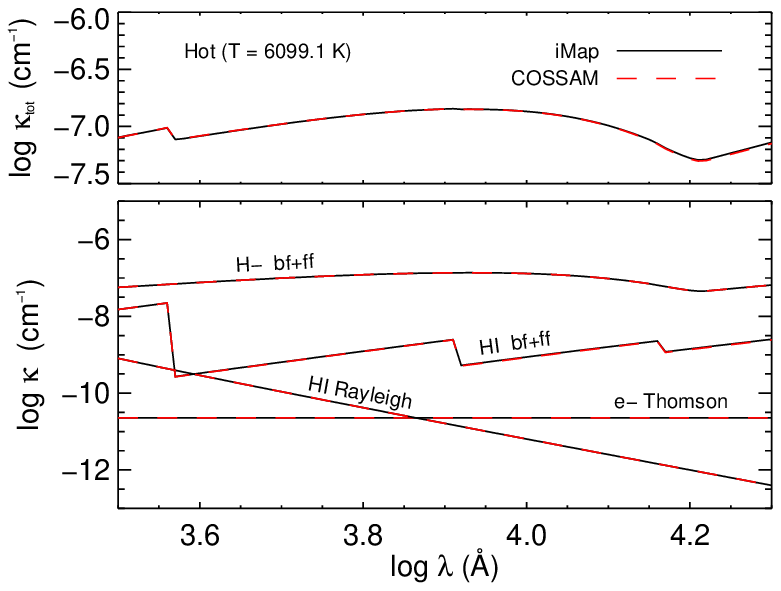}
\caption{Total continuous absorption coefficient over the wavelengths (top) calculated with \emph{iMap} (solid black line) and 
C{\tiny OSSAM} (dashed red line).
On the left side we see the calculation for the cold temperature regime and, on the right, for the hot temperature regime. 
In the bottom panel we see the absorption for the major contributors. Both codes show a very good 
agreement with rms error smaller than 1 \%.}
\label{Fig:1}
\end{figure*}

\subsection{Interagreement between iMap and C{\small OSSAM}}

For this benchmark test we follow the framework of \citet{Wade01}, who performed a detailed
analysis of three different polarized radiative transfer codes: 
C{\small OSSAM} \citep{Stift00}; I{\small NVERS}10 \citep{Pisk02}; Z{\small EEMAN}2 \citep{Land88}. 
Our goal in this section is to assess the agreement between \emph{iMap} and C{\small OSSAM} and by this indirectly 
also the agreement between \emph{iMap} and the other synthesis codes of the benchmark test of \citet{Wade01}.  

\subsubsection{Continuous opacities}

We first compare the continuous opacities and its variations over wavelength for different temperature
regimes. We therefore calculated
the continuous opacities for two different temperatures in the range of log$(\lambda)$ = 3.5 to  log$(\lambda)$ = 4.3. 
We used the Kurucz solar model atmosphere
\emph{asun} \citep{Kurucz92} to calculate the total absorption coefficient at 4017 K and 6099 K which corresponds
to a height of log$(\rho_x) = -2.185$ and log$(\rho_x) = 0.560$ respectively.
In Fig.\ref{Fig:1}, we have plotted the resulting total continuous absorption coefficients as 
calculated from both codes, C{\small OSSAM} and \emph{iMap}. Additionally, we have plotted the individual main 
contributors to the
continuum absorption (bound-free and free-free for H$^-$ and \ion{H}{i}, Rayleigh and Thompson scattering).
Both synthesis codes show a very good agreement in the continuous 
absorption coefficients as well as for the individual contributors over the entire wavelength range. 
The relative rms error is less than 1 \%.
 \begin{figure}[h]
\centering
\includegraphics[width=8cm]{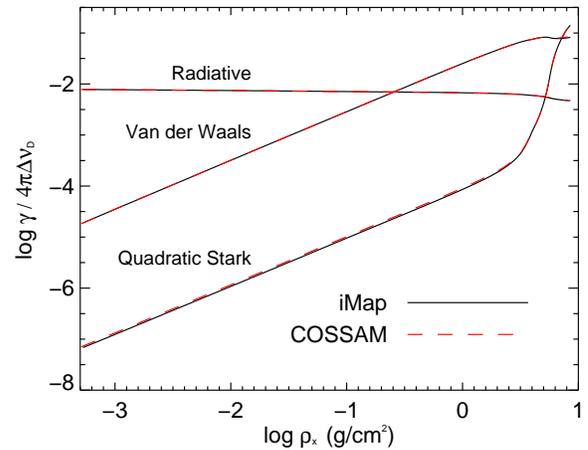}
\caption{Radiative, van-der-Waals, and quadratic Stark effect calculated with both codes under the 
\emph{asun} atmosphere of \citet{Kurucz92} (solid black \emph{iMap}, dashed red C{\tiny OSSAM}).
For all three contributions to the damping factor the two codes exhibit a very good agreement.}
\label{Fig:2}
\end{figure}
\subsubsection{Line opacities}
As line opacities are related to the total absorption and the 
individual absorption profiles, we now turn our attention to the individual damping
factors which enters into the absorption and anomalous dispersion profiles, i.e.,
Voigt- and Faraday-Voigt functions. We follow the investigation of \citet{Wade01} and
compare the radiative, van-der-Waals, and quadratic Stark damping factors. These are
calculated again within the \emph{asun} atmosphere of \citet{Kurucz92}. In Fig. \ref{Fig:2}, we see 
the run of the three damping factors with log$(\rho_x)$. We find a good agreement between the two
codes for all three damping contributors with a rms error of less then 0.1 \% for the 
radiative damping term, 0.6 \% for the van-der-Waals term, and less then 2 \% for 
the quadratic Stark effect.

\subsubsection{Local line profiles -- zero magnetic field}

\begin{figure}[t]
\centering
\includegraphics[width=8.5cm]{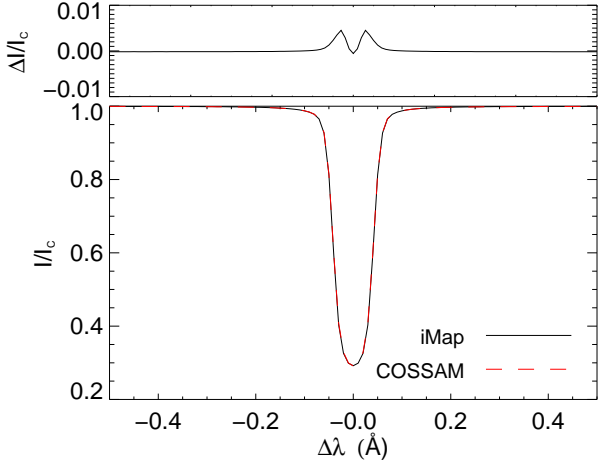}
\caption{Comparison of Stokes $I$ profiles for the zero field case (solid black \emph{iMap}, dashed red C{\tiny OSSAM}). 
Both profile calculations show a very good agreement with an rms error of less than 0.1 \% . The differences 
in the profiles are hardly recognizable in the lower plot. The top panel shows the differences $\Delta I/I_c$ 
between both profiles over the wavelength.}
\label{Fig:3}
\end{figure}

In this section, we compare the combined effect of all physical input parameters
and compare directly the resulting line profiles. For this purpose, we synthesized the
local line profile of the \ion{Fe}{i} $\lambda$ 6173 \AA\ line under the \emph{asun} atmosphere
with zero magnetic field.
The LOS is assumed to be normal to the surface 
($\theta = 0^{\circ} \rightarrow \mu = cos(\theta) = 1$) 
and furthermore we have assumed solar iron abundance of $\epsilon_{Fe} = -4.37$ 
( defined as $\log n_{Fe} / n_{tot}$ ).
In Fig. \ref{Fig:3}, we show the intensity line profile calculated with both codes (solid black is \emph{iMap}, dashed red 
is C{\small OSSAM}).
Both Stokes $I$ profiles are virtually indistinguishable within Fig. \ref{Fig:3}. The relative rms error (relative
to the continuum) between both profiles is less than 0.1 \%, a value, which is comparable to the 
inter-agreement study of \citet{Wade01}.

\subsubsection{Local line profiles -- kilo Gauss magnetic field}

\begin{figure*}[t]
\centering
\includegraphics[width=8cm]{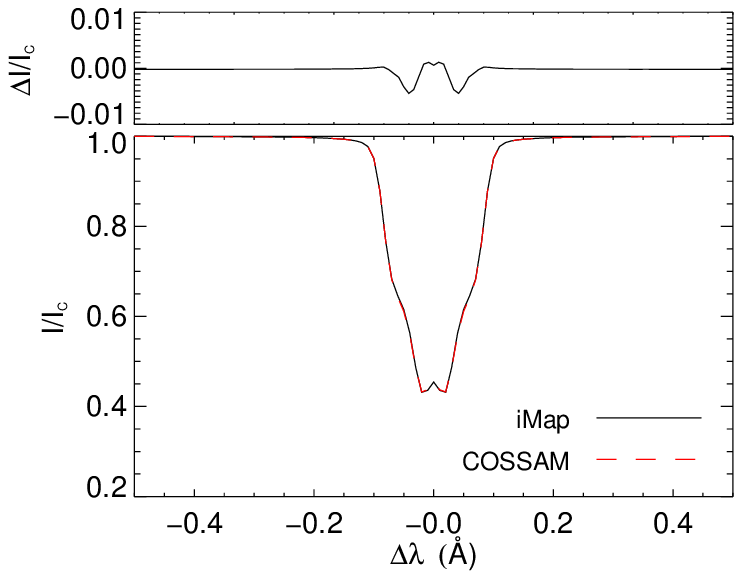}
\includegraphics[width=8cm]{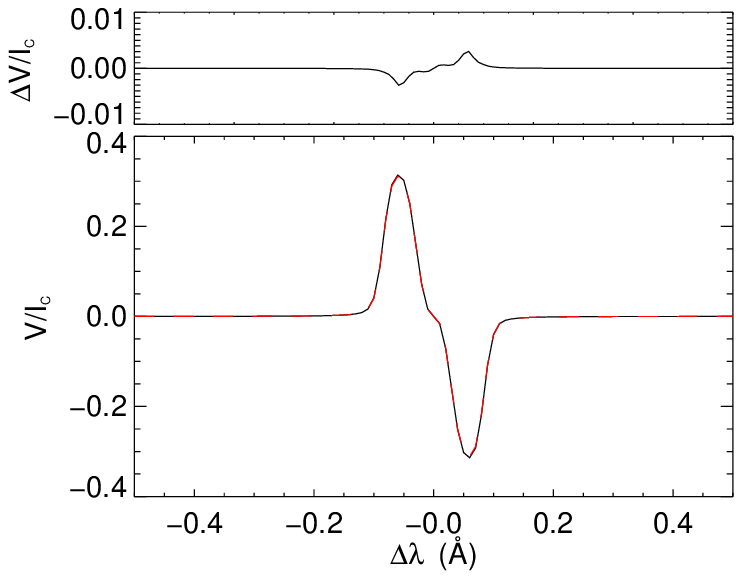}
\includegraphics[width=8cm]{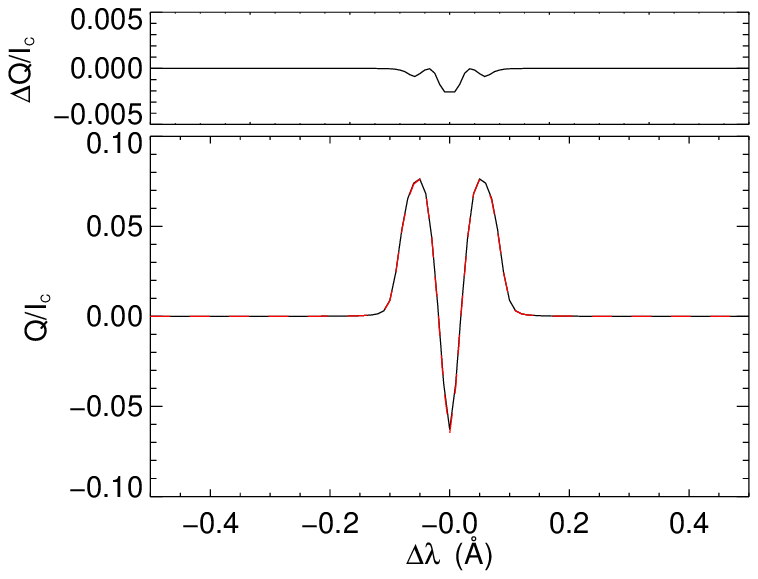}
\includegraphics[width=8cm]{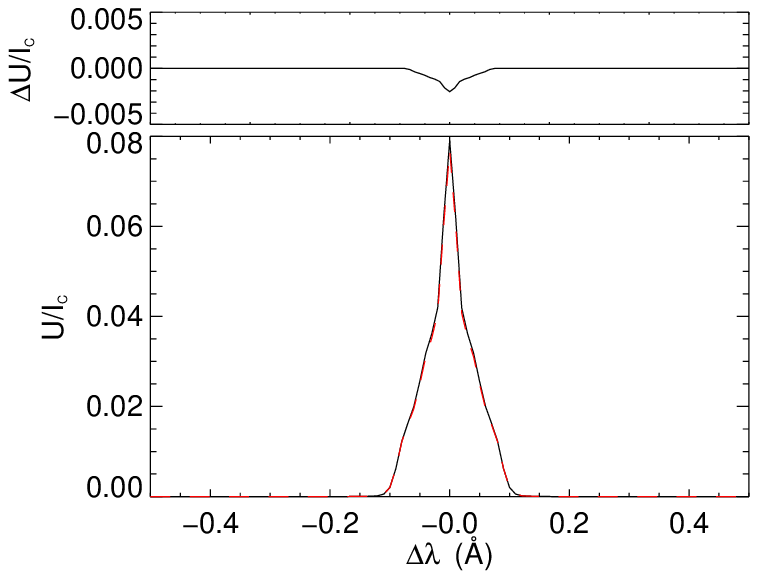}
\caption{Comparison of Stokes $I$ profiles (top, left), Stokes $V$ (top, right), Stokes $Q$ (bottom,left)
and Stokes $U$ (bottom, right) for the 1 kG case. 
Calculations made with \emph{iMap} are again in solid black lines and those made with C{\tiny OSSAM} in dashed red lines.
Difference plots are shown again on top of each profile plot.}
\label{Fig:4}
\end{figure*}

We now introduce a volume filling homogeneous (depth-independent) magnetic field of 1 kG. 
The inclination of the field (relative to the LOS) is set to 40$^{\circ}$ and the 
azimuth to 0$^{\circ}$.
All other parameters are the same as in the preceding section.
Figure \ref{Fig:4} again demonstrates the remarkable good agreement 
between the two codes for all Stokes parameter profiles. The rms errors
are as follows : for the Stokes $I$ profiles = 0.18 \%; for Stokes $V$ = 0.4 \%; 
for Stokes $Q$ = 1.0 \%; and for Stokes $U$ = 0.98 \%. The rms values are given relative to
the full amplitude \citep[see][]{Wade01} and are well within the margins of their 
interagreement study.

From these test calculations, we can already state that both synthesis codes
show a very good agreement, which is comparable to the inter-agreement between
the three synthesis codes I{\small NVERS}10, Z{\small EEMAN}2, and C{\small OSSAM}.

\section{The artificial neural network approach}
\label{Sect:3}
The basic idea of the approach is to emulate the process of 
polarized line formation by using an adaptive model, 
that is fast to evaluate, and which provides the required accuracy. 
The adaptive model we seek must have a sufficient complexity to describe 
the non-linear mapping of Eq. (\ref{Eq:1}), between the most 
prominent atmospheric input parameters and the resulting Stokes spectra.
For this purpose, we propose a supervised machine learning algorithm, e.g., 
an artificial neural network (ANN) model. 
Successful applications of ANNs in the field of
spectral analysis and Stokes profile inversions have been presented 
by \citet{Carroll01,Carroll03,Socas05b,Carroll08}.
In this section, we want to show how a popular type of ANN,
known as multilayer perceptrons (MLPs), can be used for the forward modeling approach,
i.e., the polarized spectral line synthesis.
The MLP can be regarded as a class of nonlinear function, which
performs a multivariate mapping between some input vector $\vec{x}$ and an output vector $\vec{y}$.
The network function represents a function composition of elementary
non-linear functions $g(a)$. These elementary functions are arranged in 
layers whereby each of these functions in one layer is connected via an adaptive weight vector $\vec{w}$ to
all elementary functions in the neighboring layers.
The $l-th$ output (component of the output vector $\vec{y}$) of a three layer (of weights) MLP for example 
can be written as
\begin{equation} y_{l}({\vec{x}};{\vec{w}}) =
g_{l} \left(\sum_{k=0}^{K}w_{lk}^{(3)} g_{k} \left(\sum_{j=0}^{J}w_{kj}^{(2)} g_{j}\left(\sum_{i=0}^{I}w_{ji}^{(1)}x_{i}
\right) \right) \right) \; ,
\label{Eq:6}
\end{equation}
where $x_i$ represents the $i$-th component of the input vector $\vec{x}$ and
$w_{ji}^{(1)}$ the connecting weight from the $i$-th input component to the $j$-th elementary function
$g_j$ in the first unit layer.
The weights $w_{kj}^{(2)}$ then connecting all the functions $g_j$ with the functions $g_k$ in the second unit layer 
and $w_{lk}^{(3)}$ finally connects all functions $g_k$ from the second unit layer to the functions $g_l$ in the 
third unit layer.
The capital letters (I,J,K) giving the numbers of elementary functions (units) 
in the respective unit layer.
The elementary functions $g(a)$, which are also called activation functions, are given by the
following type of sigmoid function,
\begin{equation}
g(a) = \frac{1}{1+exp(-a)} \; .
\label{Eq:7}
\end{equation} 
The network function $\vec{y}(\vec{x})$  will thus process a given input vector $\vec{x}$ 
by propagating this vector (by multiplication with the individual weight values and subsequent 
summation and evaluation of the different activation functions) through each layer of the network.

The particular function that will be implemented by the MLP is determined 
by the overall structure of the network and the individual adaptive weight values. 
The process of determining these weight values for the network function (MLP)
is called (supervised) training and is formulated as a non-linear optimization problem. 
This training is performed on the basis of a representative dataset, 
which includes the input to target relations of the underlying problem. 
This process is similar to a non-linear regression for a given
data set, but as the underlying model (i.e., MLP) is much more general, the regression function is 
not restricted to a specific predetermined (or anticipated) model.
In fact, it can be shown that a MLP provides a general framework for approximating 
arbitrary non-linear functions \citep{Bishop95}.
As for all statistical learning methods particular attention must be given to the 
dataset used for the training, which should contain a representative and statistical relevant sample of the 
underlying function we wish to approximate. In our case the input-to-target relation is dictated by our
synthesis problem and therefore the input vectors are given by the underlying atmospheric parameters 
(magnetic field, temperature etc.) and the target vectors, which reslult from the polarized radiative transfer, 
by the corresponding profiles of the Stokes spectra.
The training process (optimization) then tries to adjust the free parameters (i.e. weights) of the network
to find an approximation for the underlying generator (function) of the training data.
The optimization problem is formulated as a non-linear, least-square problem 
and is described by the following summed-square error function 
\begin{equation}
E = \frac{1}{2}\sum_{n=1}^{N}\sum_{k=1}^{C}\left\{y_{k}({\bf{x}}^{n};{\bf{w}}) -
t_{k}^{n}\right\}^{2} ,
\label{Eq:8}
\end{equation}
where N is the number of training patterns (i.e.,pairs of input-target vectors) in the
training set and C the number of output units (elements) of the network; 
$y_{k}({\vec{x}}^{n};{\vec{w}})$ the network function for the $k$-th output component 
given the $n$-th input vector $\vec{x}^n$ of the training sample. The value $t_{k}^{n}$ is the 
$k$-th (target) component of the $n$-th training vector.

The optimization of the network function 
is then performed by a non-linear least-square procedure.
Once the network is successfully trained and has converged in terms of minimizing the error 
function Eq. (\ref{Eq:8}) for the training samples, the network weights are frozen and the 
MLP, which now represents the desired approximation of the underlying function, can be 
applied to new and unknown data coming from the same parameter domain as the training data.

\subsection{Synthesis of the Stokes database and principal component analysis}

To properly train the network function Eq. (\ref{Eq:6}), we have to generate an exhaustive 
training database of synthetic Stokes spectra.
For the task of building the training database, we have calculated with our 
ZDI code \emph{iMap} a large database
of synthetic local Stokes spectra. To provide the network with the
necessary information about the underlying mapping, we have identified
the following atmospheric quantities as input parameters for the MLP: 
the temperature and pressure structure of model atmospheres (described by the effective temperature $T_{\rm{eff}}$), 
gravitation (log $g$), 
iron abundance ($\epsilon_{Fe}$), the local bulk velocity ($v_{bulk}$) of the plasma, 
microturbulence ($v_{mic}$), macroturbulence ($v_{mac}$),
magnetic field strength ($B_{loc}$), magnetic field inclination ($\gamma_{loc}$), 
magnetic field azimuth ($\chi_{loc}$) and the LOS angle between observer and the local normal ($\theta_{loc}$)
of the surface element. The latter parameter determines also the center-to-limb variation (limb darkening)
of the disk-integrated spectra. The magnetic field parameters are assumed to be depth independent.

To confine the possible combination of the input parameters,
but still provide a reasonable flexibility over a wide range of parameters for the MLP spectrum synthesis,
we divided the atmospheric quantities in a set of variable and static (fixed) parameters. The static parameters
are initially chosen and then fixed during the course of the investigation. 
The variable as well as static parameters and their associated parameter ranges are listed in Table \ref{Table:1}.

The spectral line we have synthesized is the iron line \ion{Fe}{i} $\lambda$ 6173 \AA\ ,
which exhibit a normal Zeeman triplet and is a suitable line for the Zeeman diagnostic because of its large 
effective Land\'{e} factor ($g_{\rm{eff}}$ = 2.5) and because this line experiences only minor
contribution from surrounding blends. 
For each Stokes parameter ($I$,$Q$,$U$,$V$) we have calculated 20,000 Stokes spectra by 
choosing random values for the individual input parameters from the listed parameter ranges 
(Table \ref{Table:1}). The static parameters also enter into the calculation with
the values given in Table \ref{Table:1}, but were not modified.
Using a random generator
with a uniform distribution ensures moreover that the available parameter space was
evenly sampled. Note that, even though the 5-dimensional parameter space for the variable input parameters 
is relatively sparsely sampled, MLPs and many ANNs in general are much less susceptible to what 
is described as \emph{the curse of dimensionality} (the exponential growth of hypervolume) than conventional
parameterized methods \citep{Bishop95}.

The spectral lines were calculated in a wavelength range of $\pm$ 1.0 \AA\ ,around the central wavelength, 
the spectral resolution for the synthesis was 0.01 \AA\ , which results in a total of
201 wavelength points.
Since the intrinsic dimensionality of the Stokes profiles is usually much lower than the typical wavelength
sampling, the individual line profiles can be described
in a much more compact and reduced form than in the original wavelength domain \citep{Asensio07}.
Taking advantage of the extensive training database, we have applied a principle component
analysis (PCA) to decompose the individual
Stokes profiles $\vec{x}_n$, by calculating the respective covariance matrix $\vec{C_x}$ of our training database,
\begin{equation}
\vec{C_x} = \sum_n \left ( \vec{x}_n(\lambda) - \bar{\vec{x}}(\lambda) \right ) \left ( \vec{x}_n(\lambda) - \bar{\vec{x}}(\lambda) \right )^T \; ,
\label{Eq:9}
\end{equation}
where $\lambda$ is the wavelength, $n$ the number of individual spectral line profiles used for
the analysis, 
and $\bar{\vec{x}}$ the mean Stokes profile of all spectral lines.
A new set of coordinate axes can then be determined by calculating the eigenvectors $\vec{s}_i$ and eigenvalues $\nu_i$
of the covariance matrix Eq. (\ref{Eq:9}), 
\begin{equation}
\vec{C_x} \vec{s}_i = \nu_i \vec{s}_i \; .
\label{Eq:10}
\end{equation}
In the PCA analysis, the eigenvectors $\vec{s}_i$ are grouped in
descending order according tho their eigenvalues $\nu_i$ which account for the degree of variance in the 
data (Stokes spectra).
The Stokes spectra of the database can then be described by the orthonormal set of eigenvectors,
\begin{equation}
\vec{x}_k(\lambda) = \sum_l \alpha_{k,l} \vec{s}_l(\lambda)  \; ,
\label{Eq:11}
\end{equation}
where $\alpha_{k,l} = \vec{x}_k(\lambda) \vec{s}_l(\lambda)$ is the scalar product (projection coefficient)
between the Stokes profile $\vec{x}_k(v)$ and the eigenvector $\vec{s}_l(\lambda)$. 
Retaining only the first few dominant eigenvectors (the principal components), 
which account for the majority of the variance
in the data, we find an efficient way of reducing the
dimensionality of the Stokes spectra by using only the projection coefficients of the principal components.
The representation of the $k$-th original spectra $\vec{x_k}$ can then be expressed with the first $l$ eigenvectors as
\begin{equation}
\bm{x}_k(\lambda) = \sum_{m \leq l} \alpha_{k,m} \bm{s}_m(\lambda)  \; .
\label{Eq:12}
\end{equation}
The projection coefficients $\alpha_{k,m}$ provide the reduced (decomposed) representation of the
original $k$-th profile. With this technique we have reduced the dimension of each spectra
from 201 wavelength points to 9 for Stokes $I$ and $V$
and 14 for Stokes $Q$ and $U$.
The rms reconstruction error for the PCA decomposed spectra is in all cases less then 0.1 \%.
\begin{figure*}[t!]
\centering
\includegraphics[width=8cm]{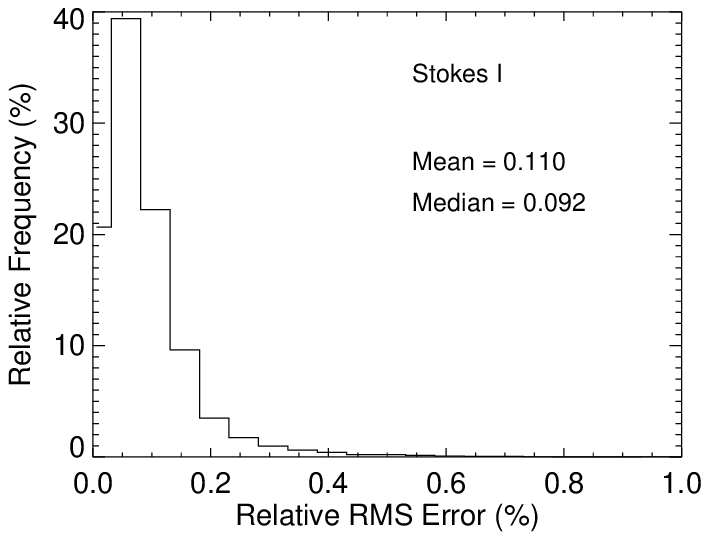}
\includegraphics[width=8cm]{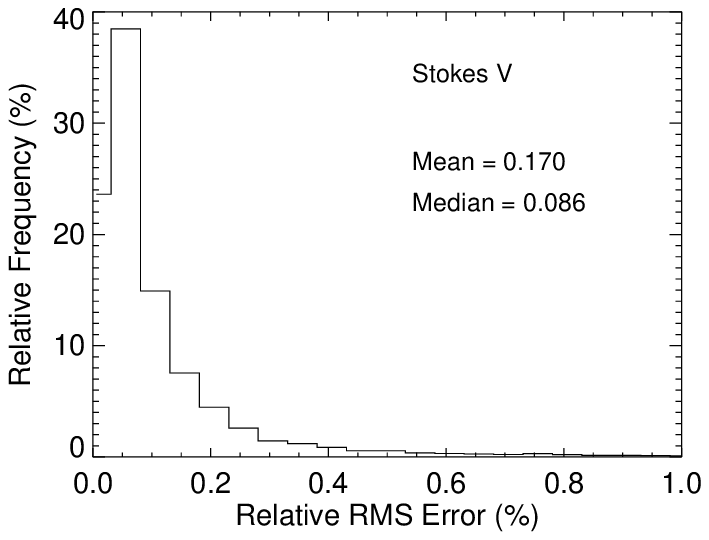}
\includegraphics[width=8cm]{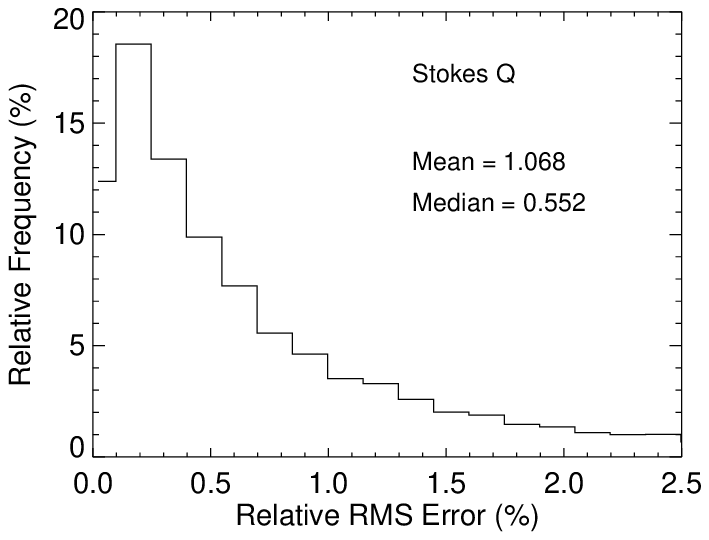}
\includegraphics[width=8cm]{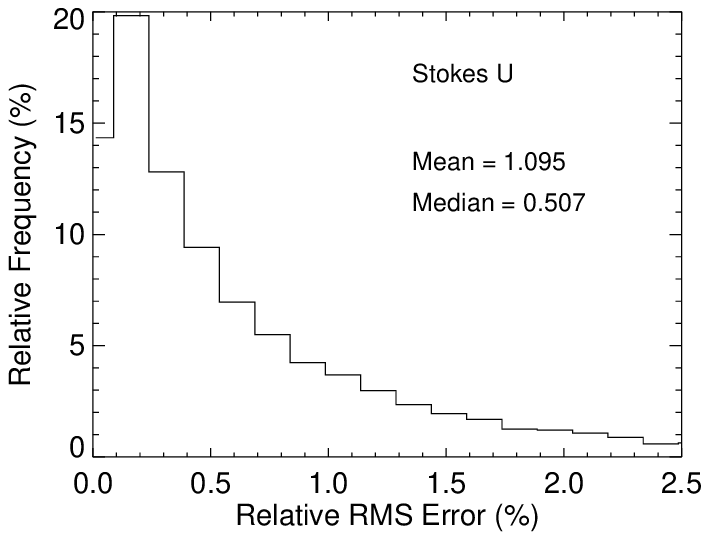}
\caption{The rms error distribution between the \emph{true} numerical- and the
MLP-calculated Stokes profiles. The validation sample consists of 13,000 profiles for each Stokes parameter.
Due to the asymmetry of the distribution, we also list the median besides the mean values.
The mean error for Stokes $I$ (top left) is 0.11 \%, 
for Stokes $V$  (top right) 0.17 \%, and 1\% for Stokes $Q$ (bottom left) 
and Stokes $U$ (bottom right). Note that the median values for Stokes $Q$ and Stokes $U$
are significantly lower than the mean values.}
\label{Fig:5}
\end{figure*}

\subsection{Training of the MLP synthesis networks}

For the task of approximating the polarized line formation process, according to Eq. (\ref{Eq:1}) 
we divided the approximation of the Stokes profiles into 5 individual
MLP network models. For each Stokes parameter ($I$,$Q$,$U$,$V$) we use a single MLP that calculates
the normalized line profiles (relative to the continuum intensity), and one MLP network that calculates the
continuum intensity on top of the atmosphere at the central wavelength of the respective spectral line.
The continuum intensity is expressed in units of 10$^6$ ergs cm$^{-2}$ s$^{-1}$ sr$^{-1}$ \AA$^{-1}$ .
Even though we could have used one MLP model
for the entire process of calculating the profiles of the complete Stokes vector, we decided to split up the
networks to avoid saturation effects during the training process,
which may be caused by the differences in the resulting magnitude of the individual Stokes 
parameter as well as by the different sensitivities (responses) to the atmospheric input parameter.
\begin{table}{t}
% title of Table
%\label{Table:1}      % is used to refer this table in the text
\centering                          % used for centering table
\begin{tabular}{c c c}        % centered columns (4 columns)
\hline\hline        % inserts double horizontal lines
lower value & variable parameter & upper value \\    % table heading
\hline                        % inserts single horizontal line
   4500 K  & $T_{\rm{eff}}$  &  6500 K   \\      % inserting body of the table
   0 G  & $B_{loc}$  &  2500 G  \\
   0$^{\circ}$  & $\gamma_{loc}$  &   180$^{\circ}$  \\ 
   0$^{\circ}$  & $\chi_{loc}$    &  180$^{\circ}$  \\ 
   0$^{\circ}$  & $\theta_{loc}$  &  90$^{\circ}$  \\  
   \hline                                  %inserts line
   Value & Static parameter \\
   \hline   
   -4.37 & $\epsilon_{Fe}$    \\
   4.0 & $log(g)$   \\
   2.0 km/s & $v_{mic}$  \\
   2.0 km/s & $v_{mac}$  \\
   0.0 km/s & $v_{bulk}$   \\     
   \hline                                   
%inserts single line   
\end{tabular}
\caption{Variable and static parameters and their corresponding parameter ranges.} 
\label{Table:1}
\end{table}

The 5 MLP network structures are implemented with three layers of weights according to Eq. (\ref{Eq:6}).
The input and output structure of the network is dictated by the problem. In our case, 
each synthesis network has as much input unit (elements) as there are variable parameters (Table \ref{Table:1}),
i.e., the effective temperature of the model atmosphere, the magnetic field strength, the inclination of the
magnetic field, azimuth of the field and the LOS between the surface normal and the observer.
The static parameters were kept fixed to their values listed in Table \ref{Table:1} and do not appear 
explicitly in the network structure.
The output structure of the network is given by the decomposed representation of the Stokes profiles.
Altogether we end up with two MLP synthesis networks (for Stokes $I$ and $V$) that have 5 input parameters and 
9 output parameters (the PCA decomposed line profiles), 
two MLP synthesis networks (for Stokes $Q$ and $U$) with 5 input and 14 output parameters 
(the PCA decomposed line profiles for $Q$ and $U$), and one MLP synthesis network for the continuum calculation, which
has 5 input parameters and one output for the continuum intensity.

Prior to the training process, the network inputs (atmospheric parameters) as well as the outputs
(decomposed spectra) of the training database were rescaled. A process that is necessary 
for the input parameters due to their different magnitudes, and also for the output parameters due to the
limited output range ([0,1]) of the sigmoid function (Eq. (\ref{Eq:7}) in the output layer
of the MLP networks, note that the input range is $(-\infty,+\infty)$.
Each individual input parameter was rescaled such that all rescaled parameters have zero mean 
and a unit standard deviation. The output values were rescaled into the linear slope [0.1,0.9] of the sigmoid function. 
The 5 synthesis networks were then trained on the basis of the synthetic database 
(20,000 Stokes profiles for each Stokes parameter) to obtain an approximate
model for the non-linear mapping between the atmospheric input parameters and the 
Stokes spectra. The effective complexity of the network and overfitting
is controlled during the training process by an early stopping technique \citep{Bishop95}. 
Overfitting is here indicated by the divergence
of the network error between the training dataset and an independent validation dataset.
For the training process and the optimization of the MLP networks we have used a 
conjugate gradient algorithm (see details \citet{Carroll01}). 
After experimentally testing with different network topologies, in terms of numbers
of hidden layers and units therein, we found a sufficiently good convergence and reduction of the squared error
for the following MLPs with three weight layers (two hidden unit layers) : Stokes $I$ and $V$  5-50-35-9, 
Stokes $Q$ and $U$ 5-55-55-14.  

After a successful convergence of the training process, the network weights of the 
individual MLPs are frozen and the networks now (in application mode) represent the
desired approximation of the spectrum synthesis function for the iron line \ion{Fe}{i} $\lambda$ 6173 \AA\ .

\section{Results}
\label{Sect:4}

To evaluate the performance of the trained MLPs, we have created another
independent validation database of 13,000 synthetic local Stokes profiles for 
each Stokes parameter (52,000 in total) with our \emph{iMap} code. The combinations 
of the input parameters are again determined by a random generator from the parameter ranges
listed in Table \ref{Table:1}. The calculated Stokes spectra are then decomposed 
by the PCA technique as described above. To ensure
a valid decomposition, the new validation spectra are decomposed on the basis of the training database.
Also, for the final input scaling process ,the mean and standard deviations from the training database are used.  
To assess the accuracy of the MLP calculation we compare the results of the MLP network
calculation with the \emph{real} Stokes profiles numerically calculated with \emph{iMap}
using the DELO integration method.
Prior to the comparison, the calculated (scaled) PCA profiles from the MLP were rescaled and converted (reconstructed) to
the wavelength domain.

\subsection{Local Stokes profiles}
\label{Local}
\begin{figure}[t!]
\centering
\includegraphics[width=8.5cm]{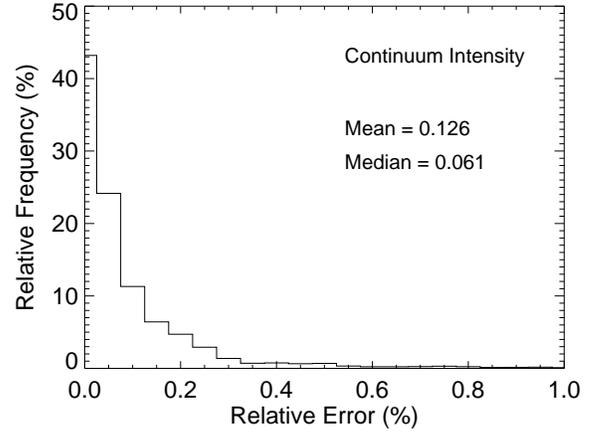}
\caption{The distribution of the relative discrepancy between the numerically-calculated
continuum intensity and the MLP calculation. The mean rms error is 0.126 \% and the median  
0.061 \% which show that the MLP is able to approximate the calculation of the continuum intensity
with high accuracy.}
\label{Fig:6}
\end{figure}

The impressive results for the 52,000 local test Stokes spectra of the validation set are illustrated
in the histograms in Fig. \ref{Fig:5}.
The relative rms error is again calculated relative to the continuum intensity for Stokes $I$ and
to the full amplitude for Stokes $V$,$Q$, and $U$.
The mean rms error for the Stokes $I$ profile calculation with the MLP is as low as 0.11 \%;
for Stokes $V$, 0.17 \%; and for Stokes $Q$ and $U$, slightly above 1 \%. Note that due to the pronounced asymmetry
of the Stokes $Q$ and $U$ error distribution the median of the distribution is only 0.5 \% .
The reason why the synthesis of the Stokes $Q$ and Stokes $U$ profiles are systematically larger than those
for Stokes $I$ and $V$ is caused mainly by two effects, first,
for relatively weak fields (and thus small profiles) Stokes $Q$ and Stokes $U$ are second order 
effects \citep{Landi92} which results in increased relative error contribution, and second,
both components periodically drop to zero for certain
azimuthal and inclination values, which again leads to very small resulting profiles and hence to an 
increase in the relative error (see also discussion below).  
Figure \ref{Fig:6} shows the distribution of the relative error compared with the numerical
calculated value of the continuum intensity with \emph{iMap}, which also show a very good agreement.

Even though effects like the temperature variation and magnetic field variation
leave their characteristic imprint in the Stokes spectra, such that an
ANN is able to approximate this behavior, it should be kept in mind
however that this will be done in a complex and non-linear fashion by the transport of polarized radiation. 
In this way, the approximation capabilities and accuracy of the trained MLPs are quite remarkable.
The MLPs have clearly been able to disentangle the effects of the different variable input parameter to
calculate the correct Stokes spectra.
Note that for the Stokes profiles as well as for the continuum synthesis,
the effects of the LOS angle $\theta$, which leads to a variation of the temperature and pressure stratification
over the optical depth scale, and which is responsible for the limb darkening effect, could apparently be well 
distinguished by the synthesis networks from other competing effects like the magnetic field strength, 
inclination, and the effective temperature of the model atmosphere. 

\begin{figure*}[t!]
\centering
\includegraphics[width=4.5cm]{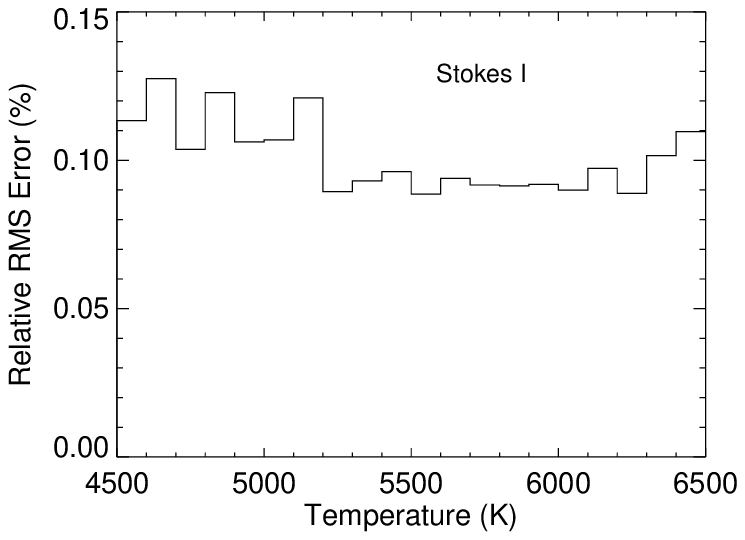}
\includegraphics[width=4.5cm]{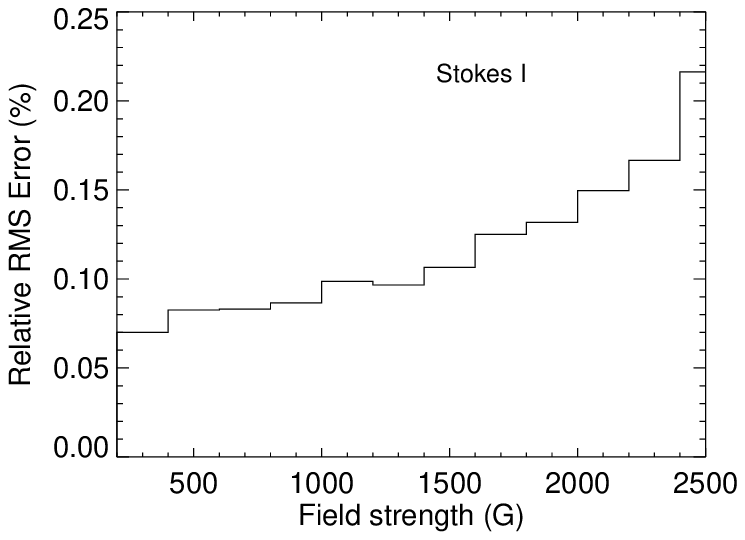}
\includegraphics[width=4.5cm]{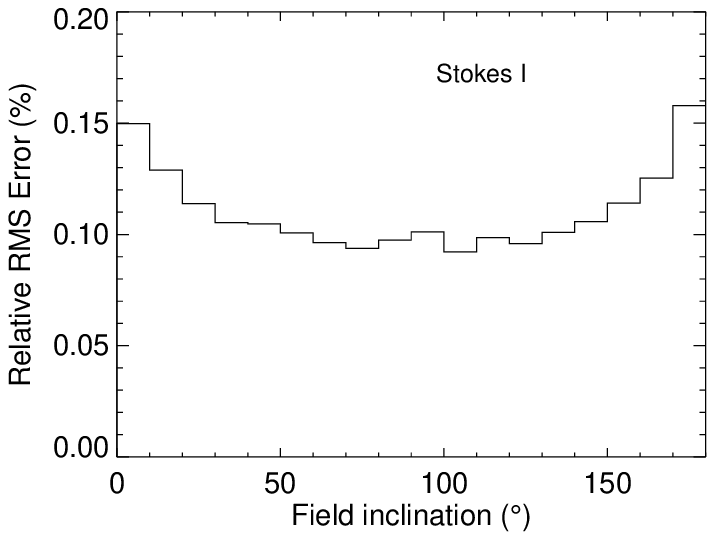}
\includegraphics[width=4.5cm]{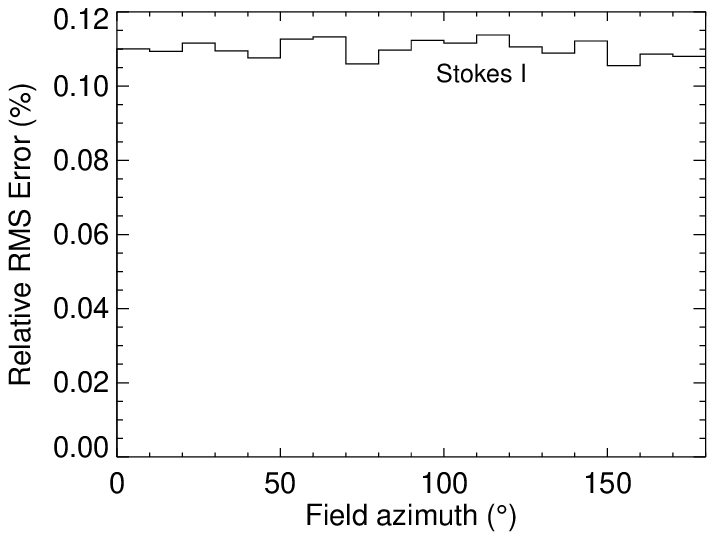}
\includegraphics[width=4.5cm]{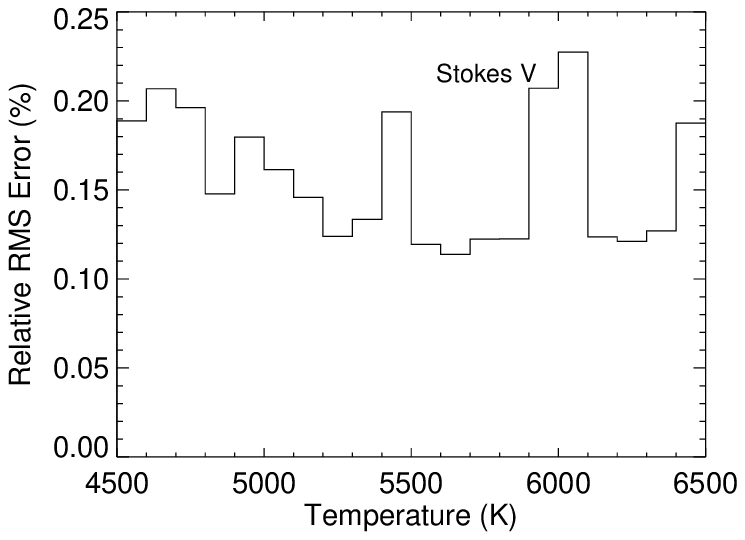}
\includegraphics[width=4.5cm]{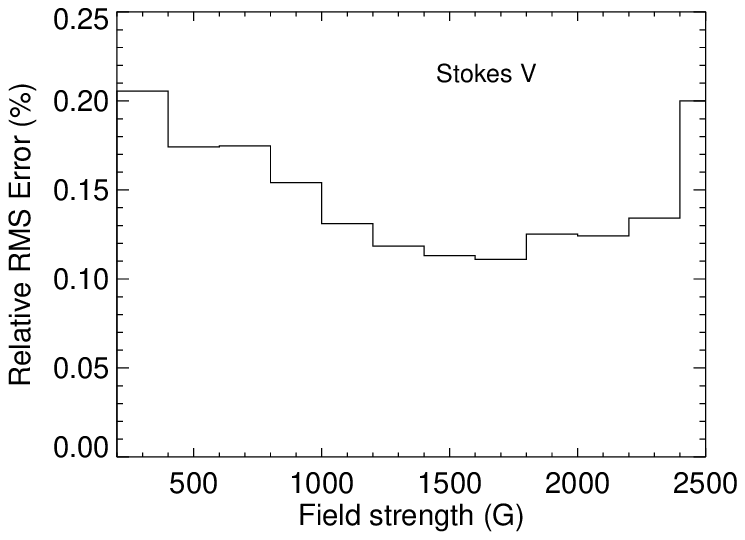}
\includegraphics[width=4.5cm]{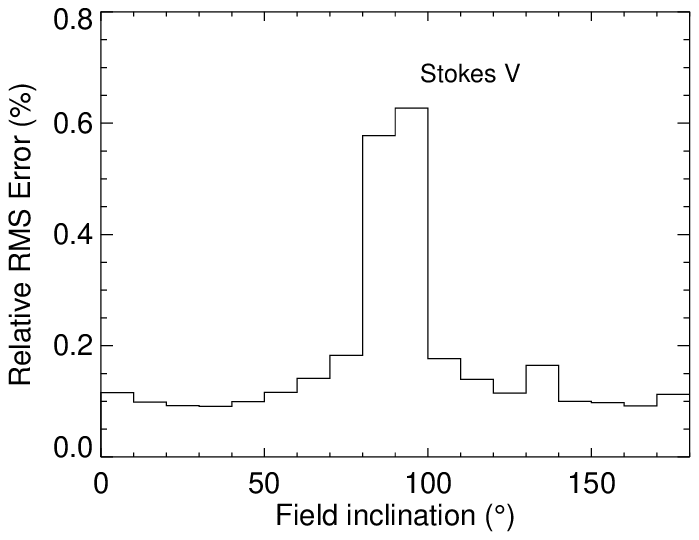}
\includegraphics[width=4.5cm]{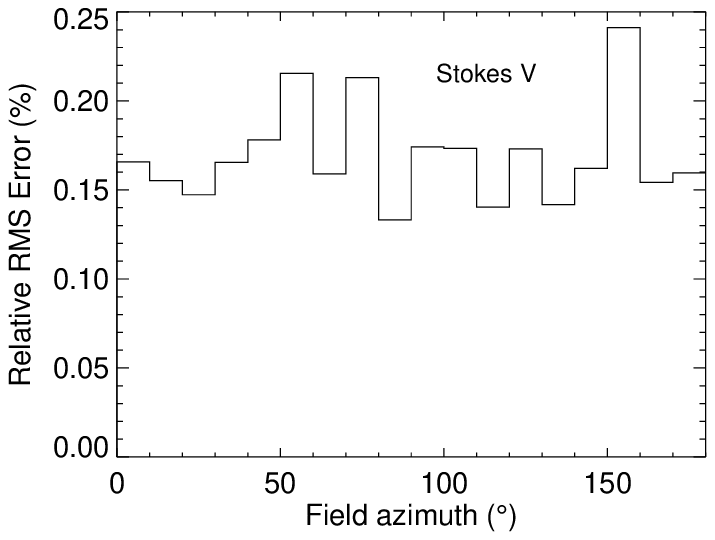}
\includegraphics[width=4.5cm]{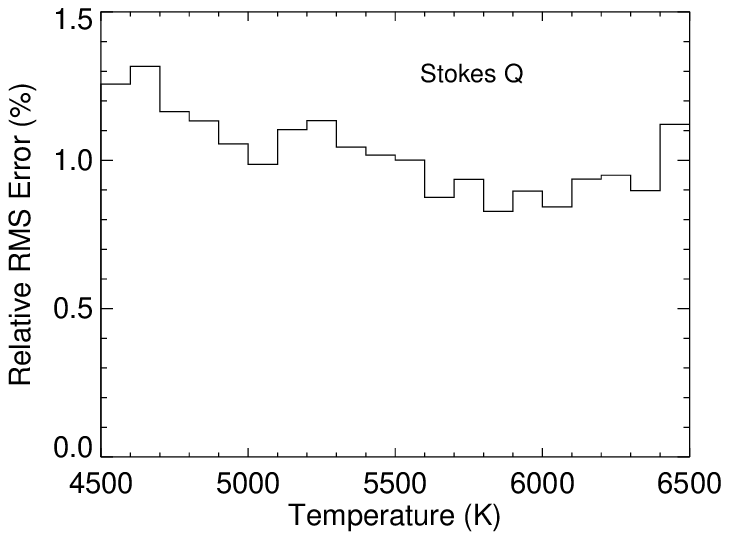}
\includegraphics[width=4.5cm]{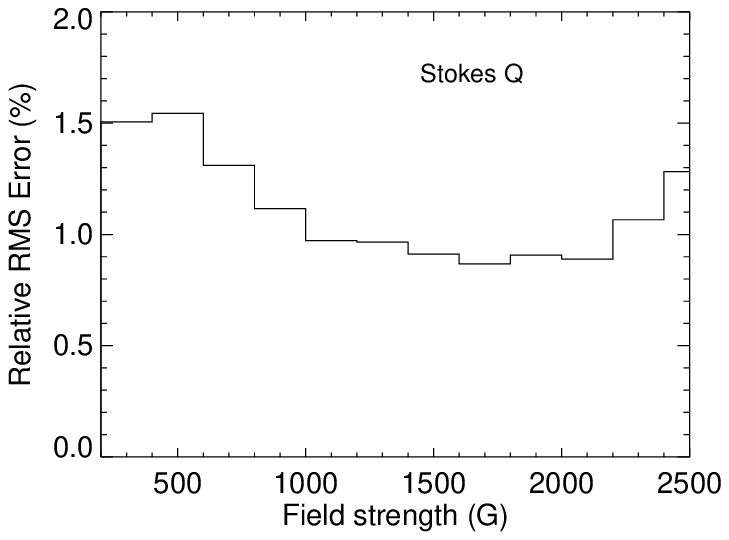}
\includegraphics[width=4.5cm]{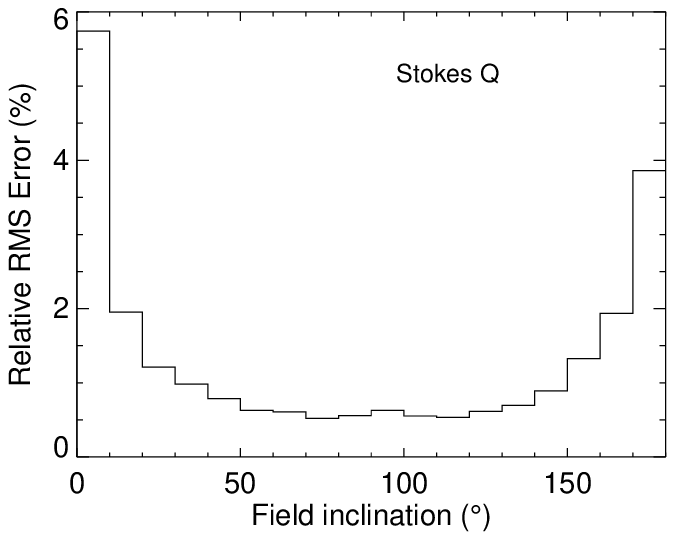}
\includegraphics[width=4.5cm]{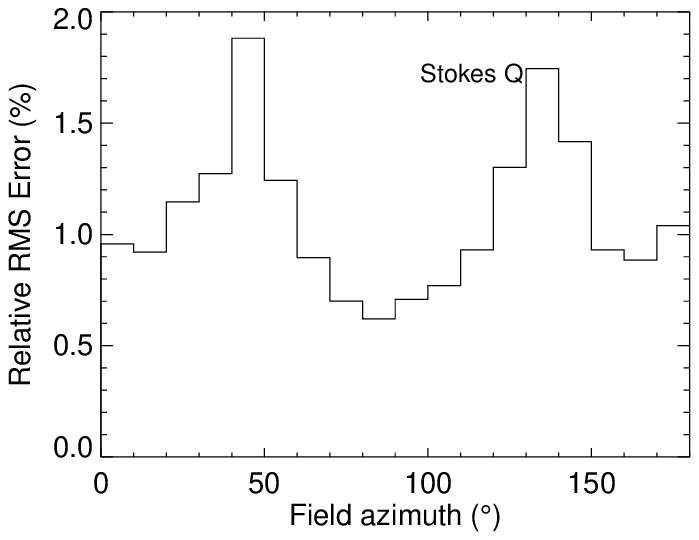}
\includegraphics[width=4.5cm]{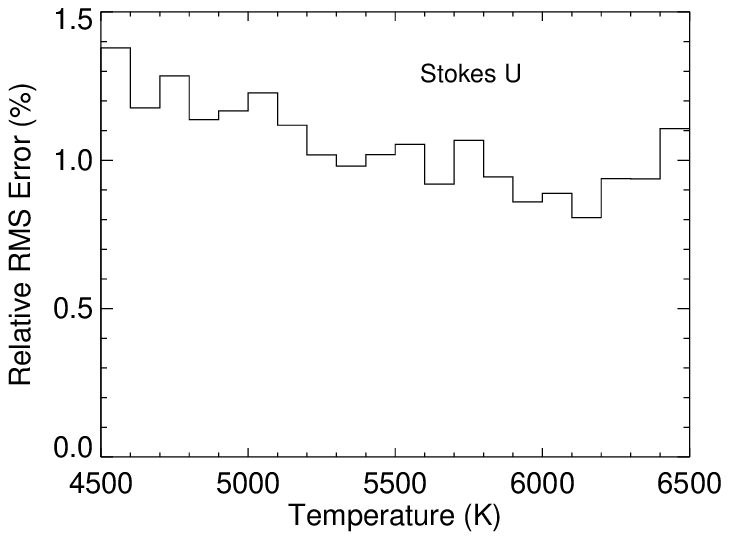}
\includegraphics[width=4.5cm]{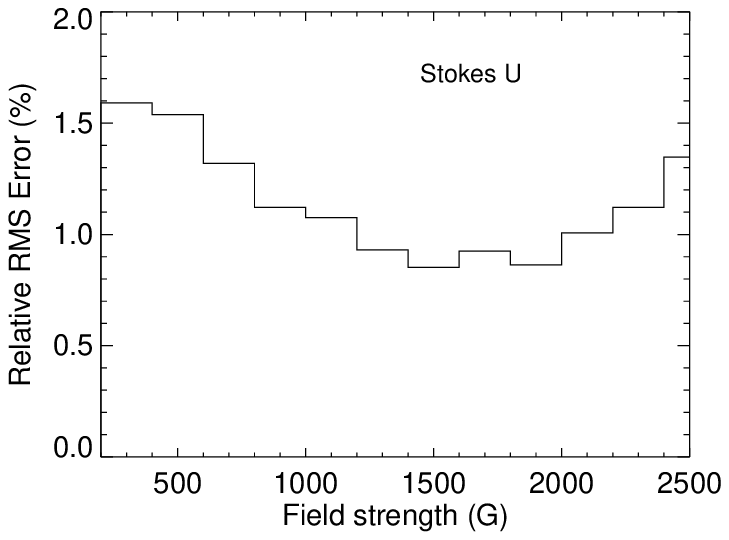}
\includegraphics[width=4.5cm]{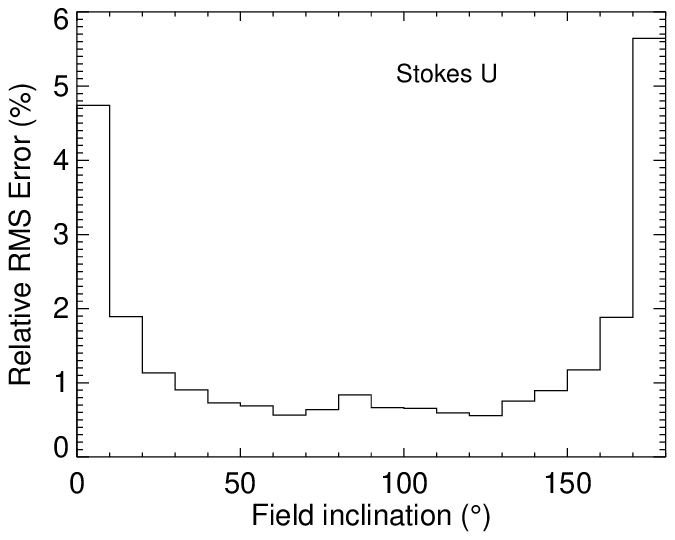}
\includegraphics[width=4.5cm]{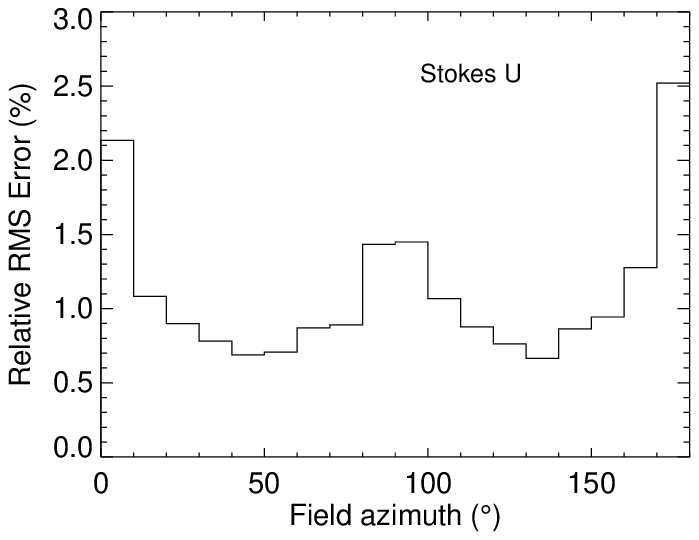}
\caption{The distribution of the rms errors for individual atmospheric parameters. 
In each row, the distributions for one Stokes parameter are listed: 
(Stokes $I$, first row; Stokes $V$, second row; Stokes $Q$, third row; Stokes $U$, fourth row). 
In each column, the distributions for a particular atmospheric parameters are given: 
temperature (1st column), field strength (2nd column),
field inclination (3rd column), and the field azimuth (4th column).}
\label{Fig:7}
\end{figure*}

To display how the errors are distributed over the main input parameter, we show in Fig. \ref{Fig:7}
the relative rms errors over four atmospheric parameters.
For the Stokes $I$ profiles (Fig. \ref{Fig:7}, top row) we obtain a relative homogeneous distribution
for the effective temperature (top row, first column) as well as for the magnetic field azimuth (top row, fourth column) which,
of course, plays hardly any role for the Stokes $I$ profile.
The situation is different for the magnetic field inclination (top row third column), 
although the rms error is 
quite low throughout the  field inclination regime (0$^{\circ}$ to 180$^{\circ}$) one notice the
increase toward the longitudinal orientation of the field 
(close to 0$^{\circ}$ and 180$^{\circ}$). This can be explained by the quasi bias (border effects) in the 
uniformly distributed training database where the pure longitudinal effect with a 
vanishing pi-component is underrepresented in the database and the more common 
non-longitudinal effect (elliptical polarization) with three contributions (two sigma- and one 
pi-components) is far more common. Nevertheless, note the remarkable agreement through the entire
range.
For the Stokes $I$ error over the magnetic field strength (top row, second column) we get a somewhat counterintuitive
result, which shows an increase of the error with field strength. One would assume a decrease
of the error since for increasing field strength the individual Zeeman components are increasingly 
separated. For strong enough fields, such that the separation of the individual Zeeman 
components is well beyond the spectral line width, one would assume that the linear dependence
of the Zeeman separation upon the field strength should be recognizable for the MLP and
therefore the functional behavior should be even easier to \emph{learn} for the MLP. 
This is in principle true but the PCA decomposition of the spectra offsets this effect and
acts in the opposite direction. 
The decomposition of a completely Zeeman-splitted Stokes $I$ profile on the basis of the current dataset 
requires a stronger representation of the higher-order (i.e., less significant) eigenprofiles 
to describe the entire profile correctly.
This means that the projection coefficients corresponding to these higher-order eigenprofiles have a significant 
value. Despite the simpler functional dependence in the Zeeman saturated regime, the PCA decomposed Stokes vector 
gets more complicated due to the non-negligible, higher-order projection coefficients. 
This effect could be easily diminished by using different datasets for the weak-field
and the strong Zeeman saturated regime.
However, note that the maximum error for strong fields is still remarkable low (lower than 0.22 \%) 
and we therefore refrain from splitting the database in this study.

For the Stokes $V$ profiles (Fig. \ref{Fig:7}, second row) we also get very satisfying results. Over the entire 
temperature range (second row, first column) we see a low error, which is around the total mean value of 0.17 \%. 
The slight peak around 5500 and 6000 K can be
explained by the coincidental accumulation of low inclination angles in that region.
For the rms error over the field strength (second row, second column) we notice a slight systematic increase for smaller
field strength. This can be explained by the smaller amplitude
of the Stokes $V$ profiles in that range which are harder to disentangle from other
effects like temperature and inclination. For higher field strengths, the Stokes $V$ profiles 
get increasingly separated, which again affects the PCA representation of these profiles 
and thus the reconstruction error (see above).
The overall error for the magnetic field strength is also very close to the total mean of 0.17 \%. 
As the Stokes $V$ signals are almost independent
of the azimuthal angle (except for small magneto-optical effects), we see no significant variation
along the azimuthal angle (second row, fourth column). 
For the inclination (second row, third column), we see a rise in the error for angles around 90$^{\circ}$ which 
displays the $cos(\gamma$) dependence of the Stokes $V$ profile on the inclination angle (Eq. \ref{Eq:5}), 
which leads to a rapid decrease to around 90$^{\circ}$. 
Although the absolute error is not much affected by the decrease of the
signal to around 90$^{\circ}$ ,the relative error increases for very small signals.
However, note that the rms error is even for the very weak signals around 90$^{\circ}$ lower than 0.7 \%.

For Stokes $Q$ and $U$, we obtain similar good results (Fig. \ref{Fig:7} third and fourth row). In both cases, we see 
a relative homogeneous error distribution over the temperature (first column) as well as for the field strength (second column).
The error distribution over the inclination (third column) shows a relatively strong increase toward 0$^{\circ}$
and 180$^{\circ}$. This increase can be understood by the smallness of the resultant Stokes $Q$ and $U$ profiles
for these inclination angles and the associated increase in the relative error.
But note that even for strongly inclined fields the MLP synthesis yields a rms error that is smaller than 6 \%.
The distribution reflects here the $cos(2 \chi$) and $sin(2 \chi$) dependence (see Eq. \ref{Eq:5}) on the 
azimuthal angle, which leads to a rapid decrease of the Stokes signal 
at these angles and therefore again to an increase of the relative
error. These two effects in conjunction (inclination and azimuth) are the main causes for the 
larger relative error of the Stokes $Q$ and $U$ synthesis networks compared to the Stokes $I$ and $V$
networks.  
But nevertheless both synthesis networks (Stokes $Q$ and $U$)  have \emph{learned} to approximate
the complicated relation and perform remarkably well 
over the entire range of parameters.

\begin{figure*}[t!]
\centering
\includegraphics[width=4.3cm]{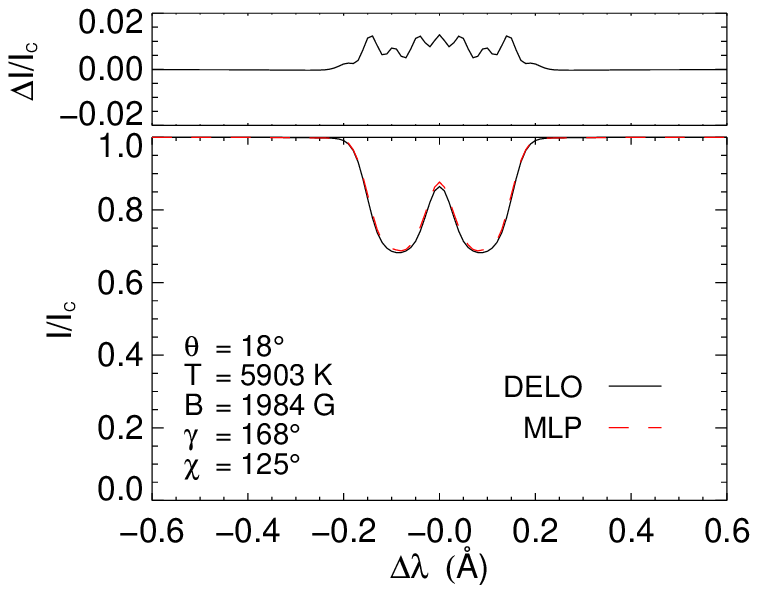}
\includegraphics[width=4.3cm]{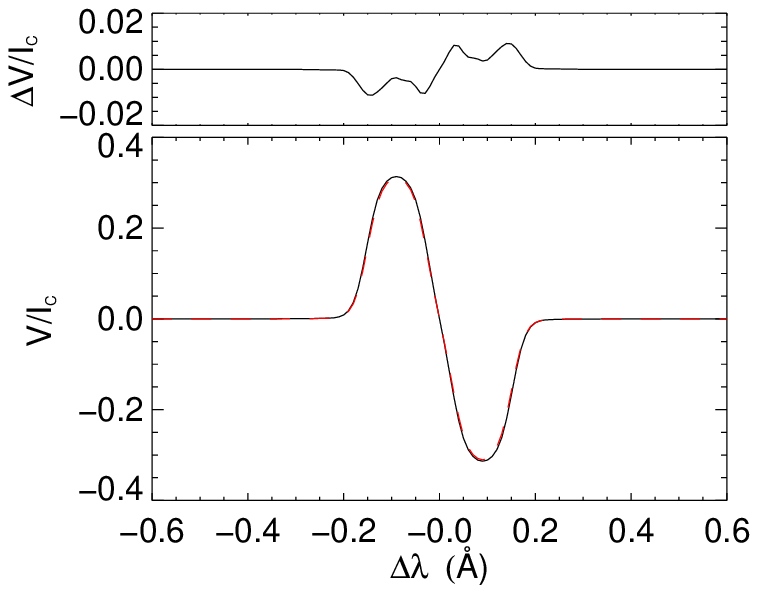}
\includegraphics[width=4.3cm]{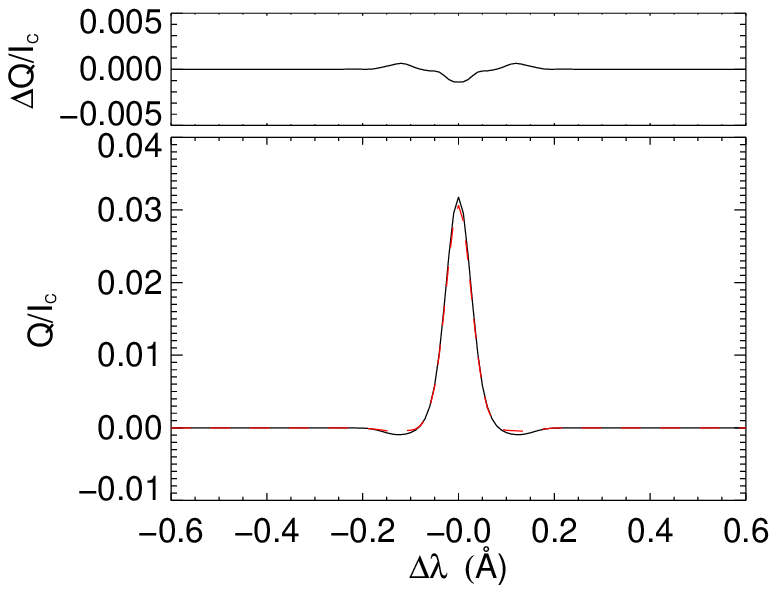}
\includegraphics[width=4.3cm]{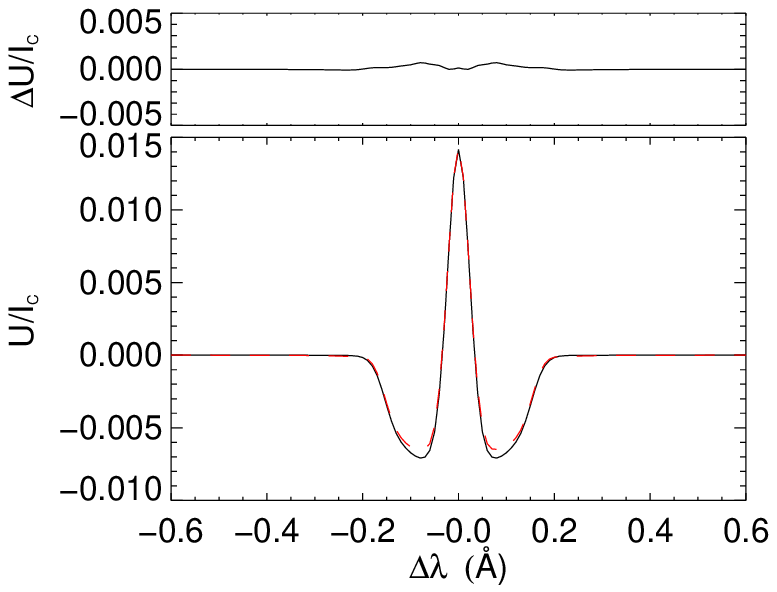}
\includegraphics[width=4.3cm]{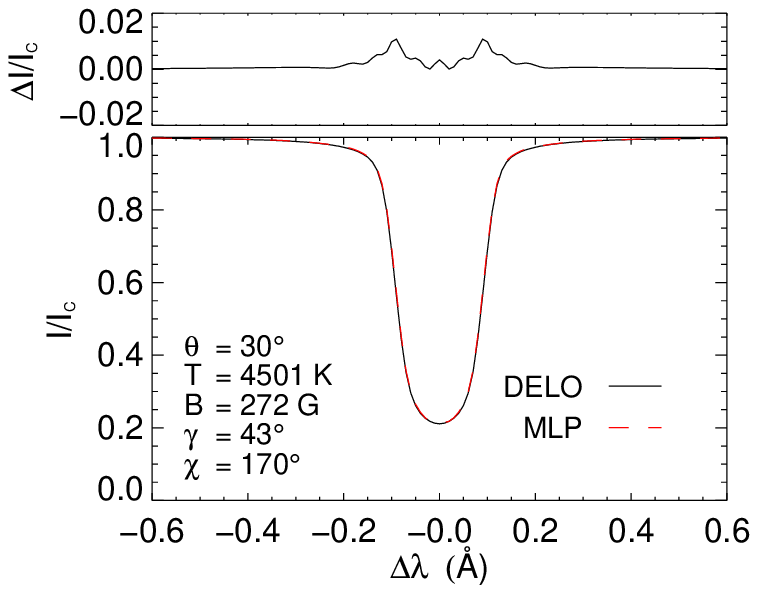}
\includegraphics[width=4.3cm]{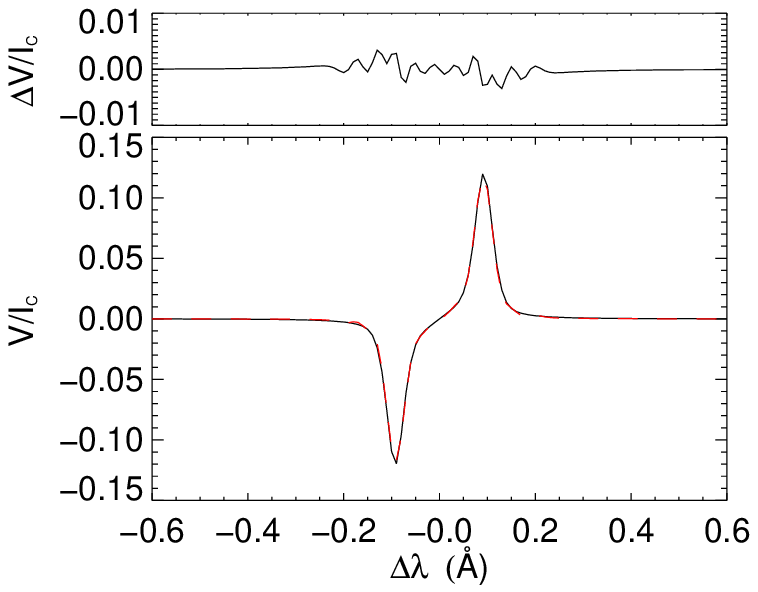}
\includegraphics[width=4.3cm]{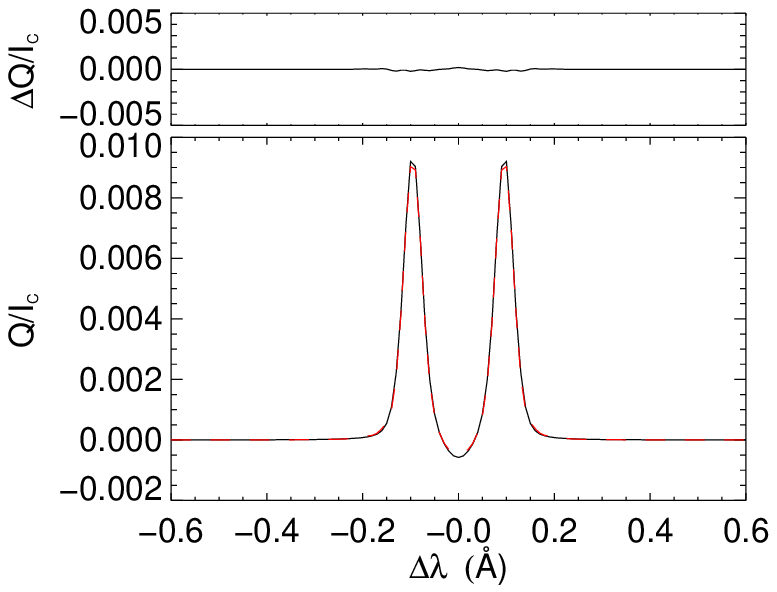}
\includegraphics[width=4.3cm]{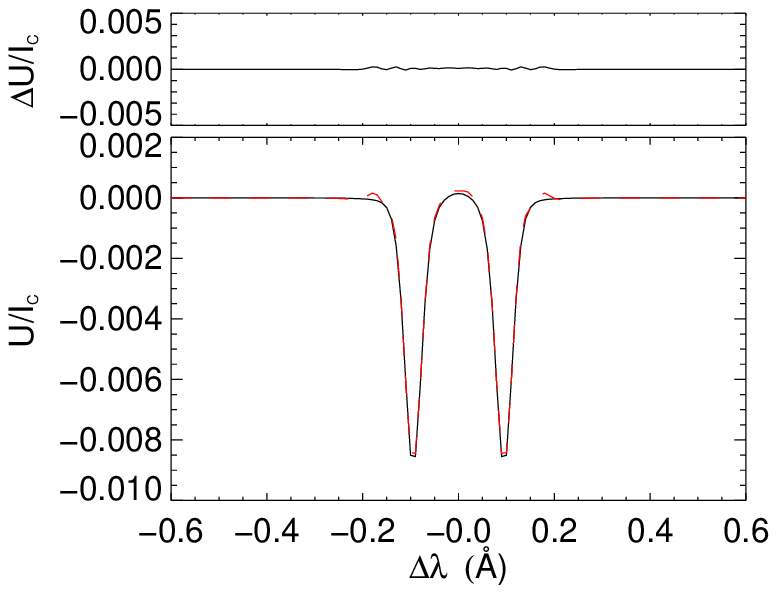}
\includegraphics[width=4.3cm]{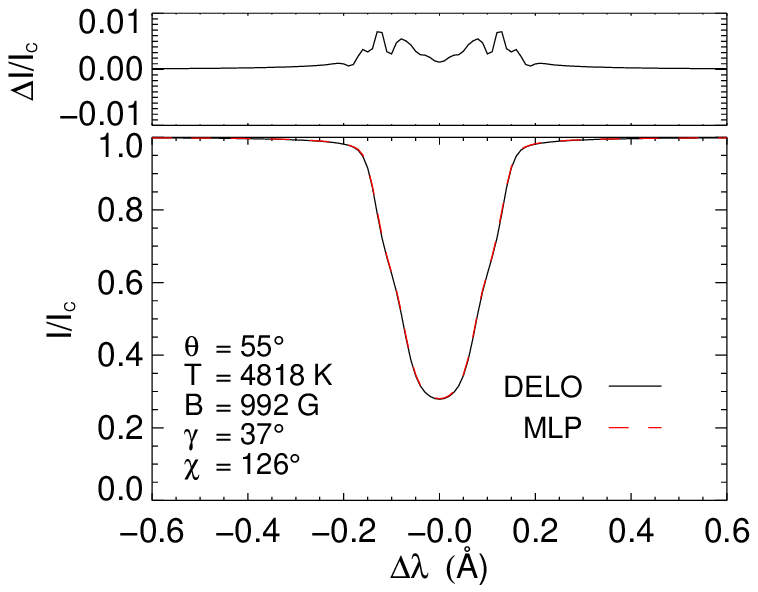}
\includegraphics[width=4.3cm]{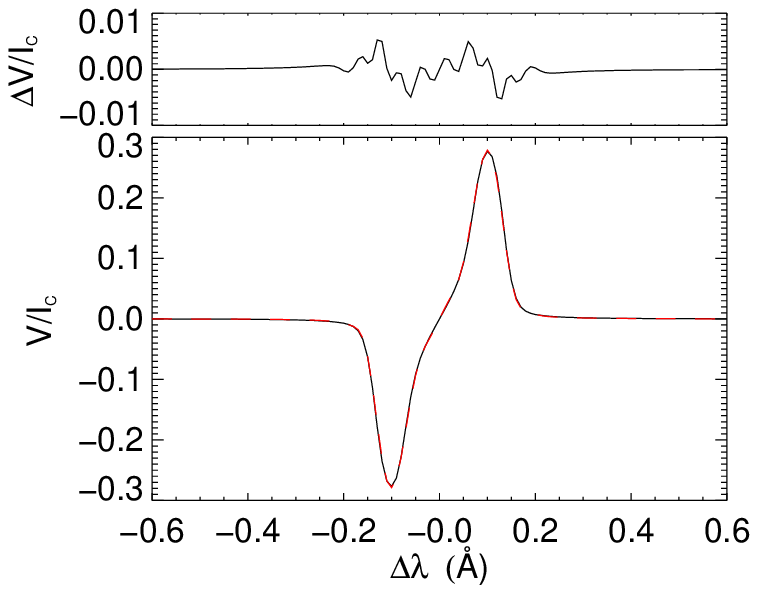}
\includegraphics[width=4.3cm]{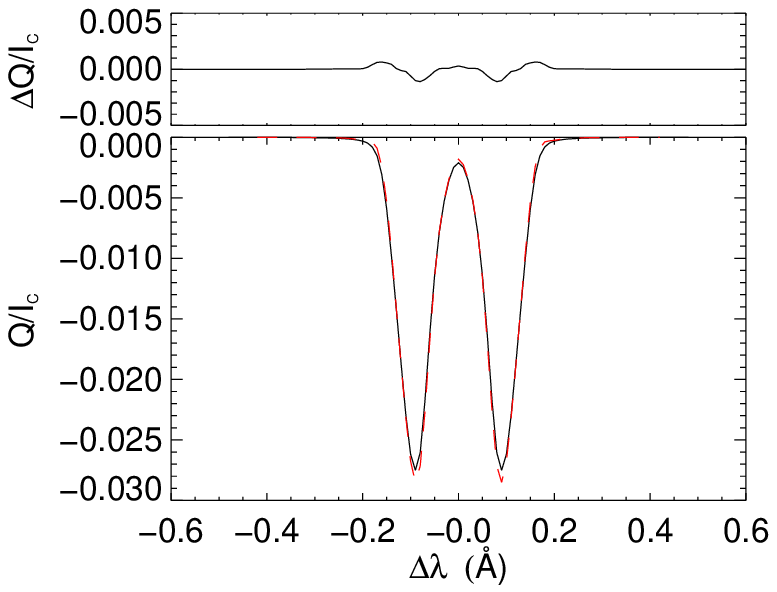}
\includegraphics[width=4.3cm]{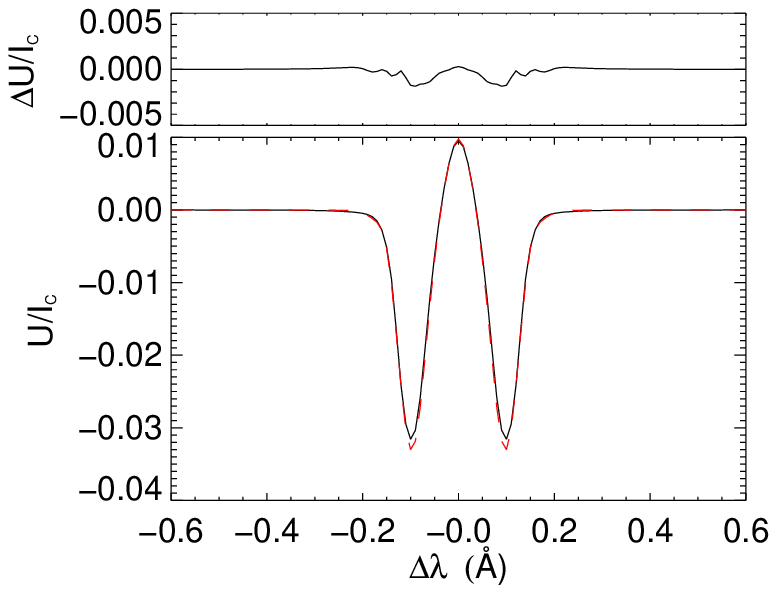}
\includegraphics[width=4.3cm]{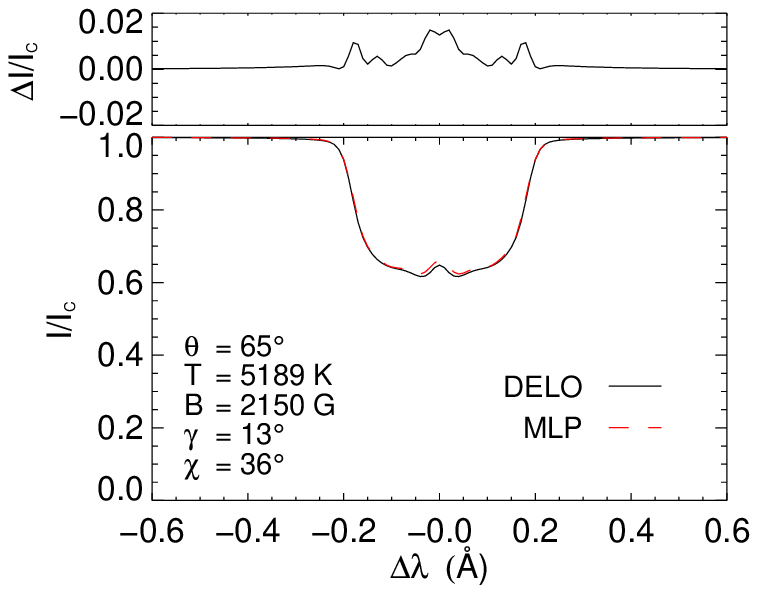}
\includegraphics[width=4.3cm]{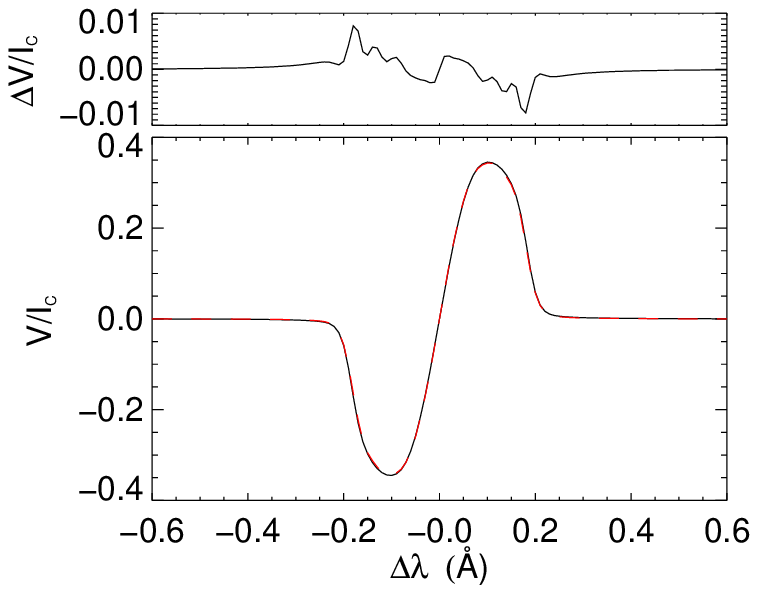}
\includegraphics[width=4.3cm]{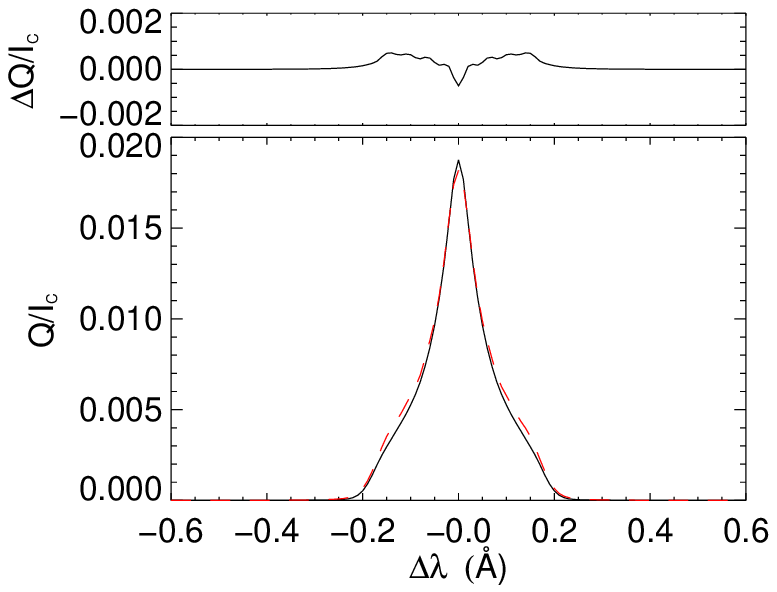}
\includegraphics[width=4.3cm]{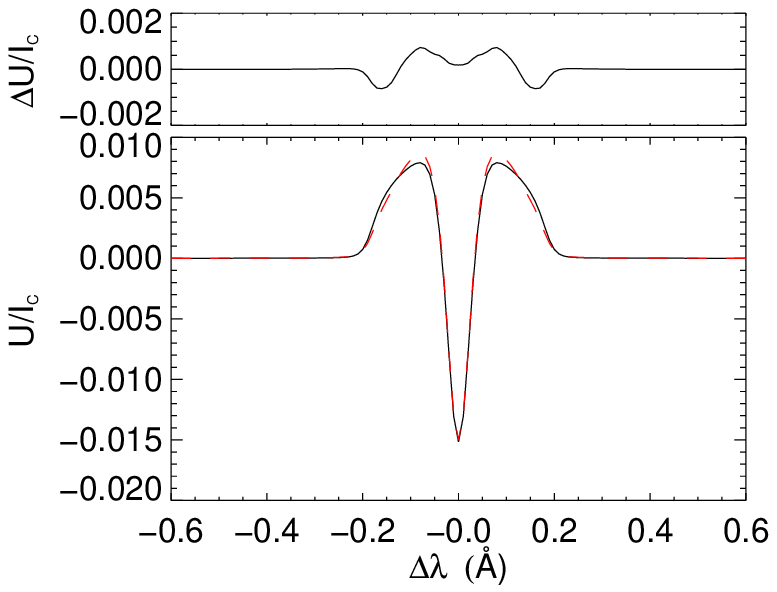}
\includegraphics[width=4.3cm]{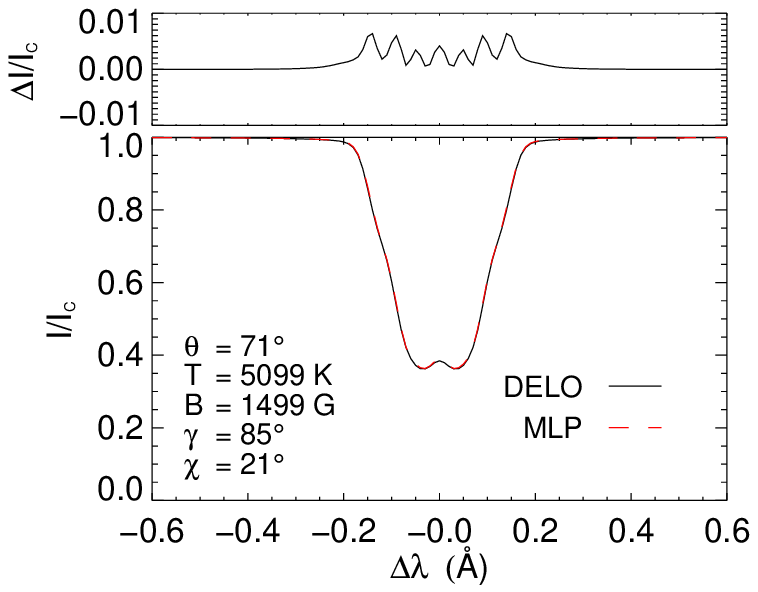}
\includegraphics[width=4.3cm]{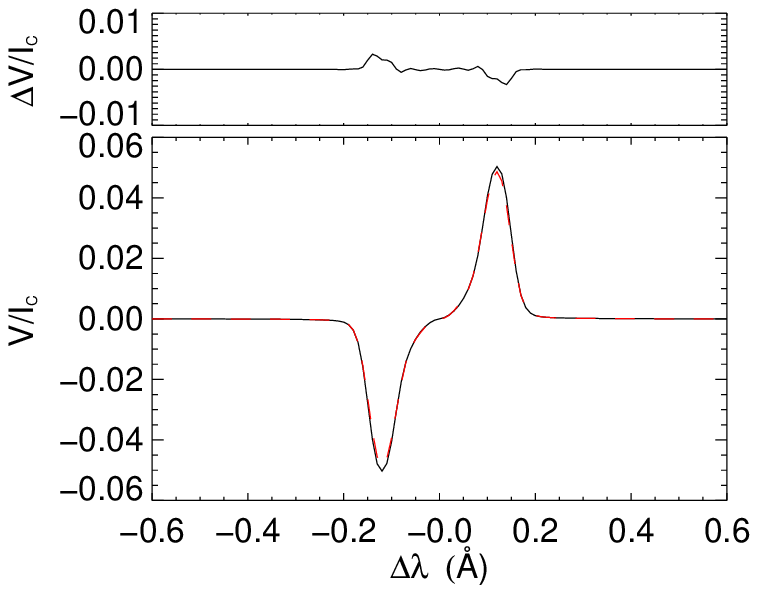}
\includegraphics[width=4.3cm]{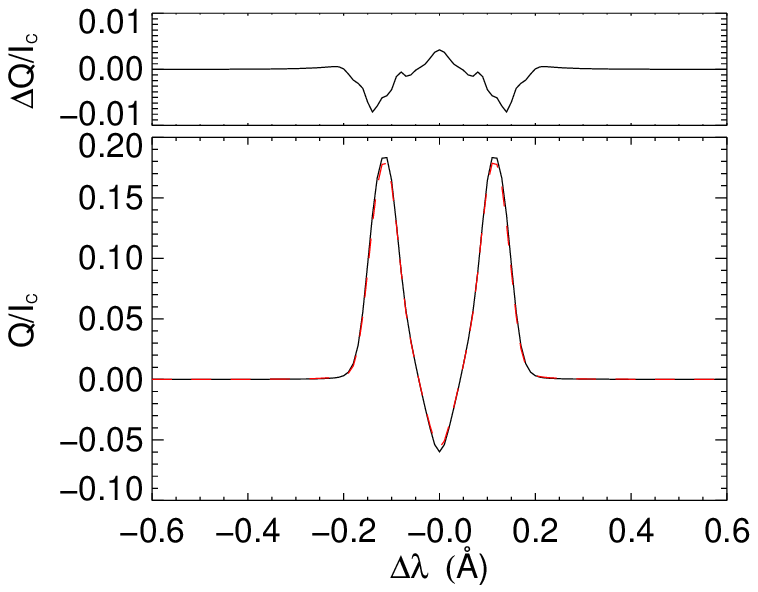}
\includegraphics[width=4.3cm]{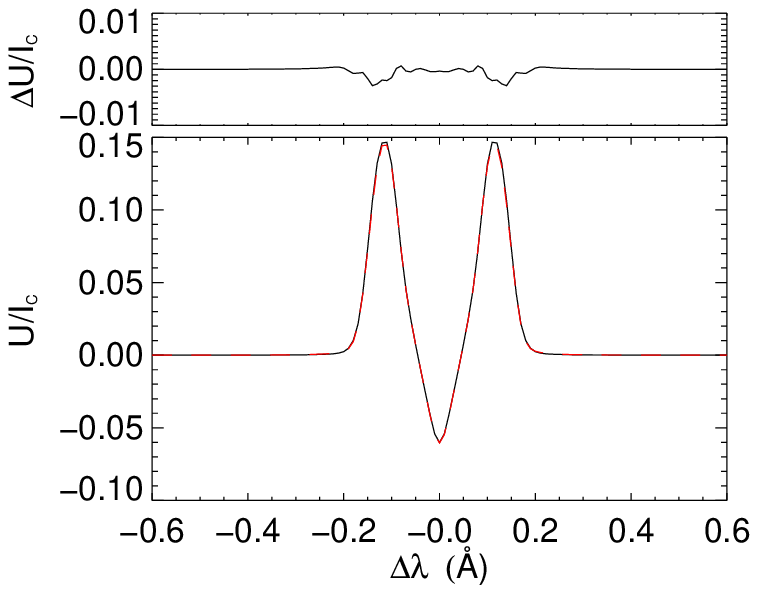}
\includegraphics[width=4.3cm]{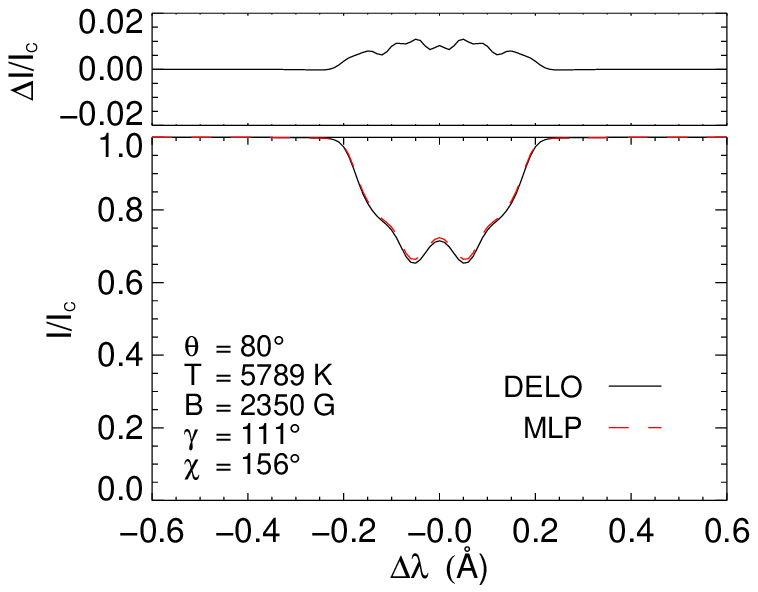}
\includegraphics[width=4.3cm]{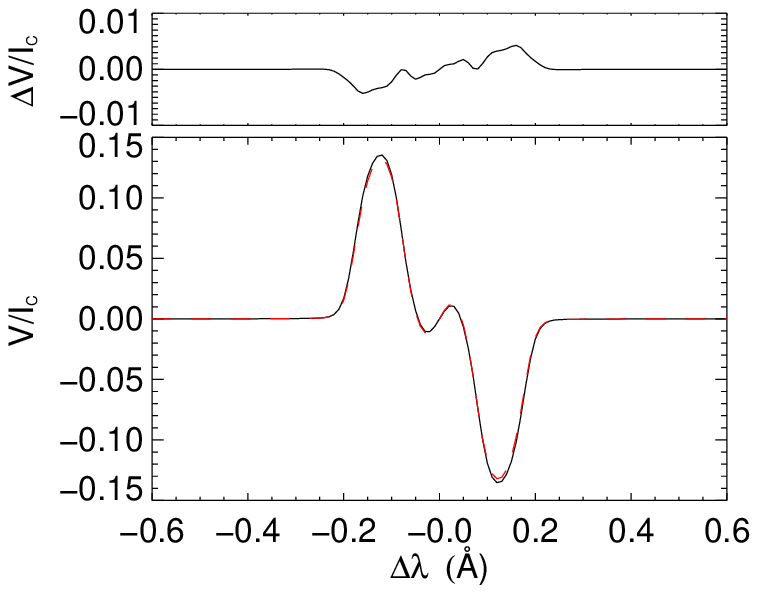}
\includegraphics[width=4.3cm]{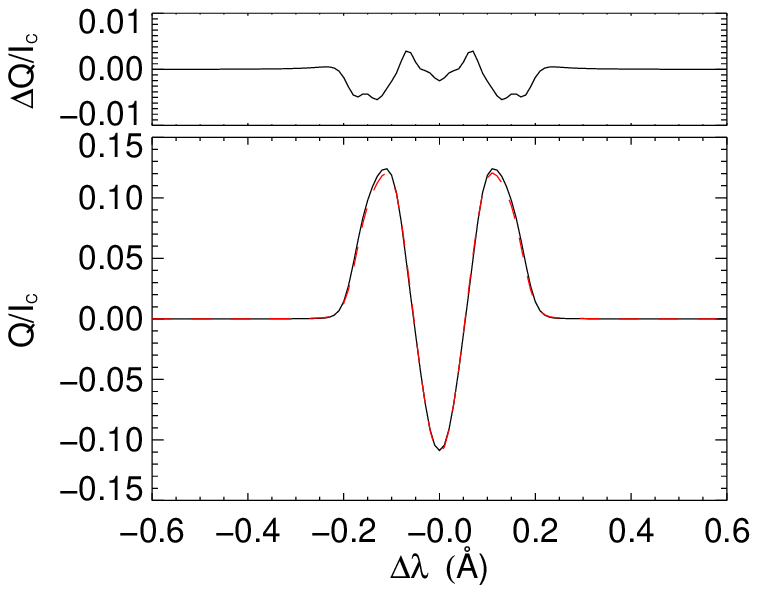}
\includegraphics[width=4.3cm]{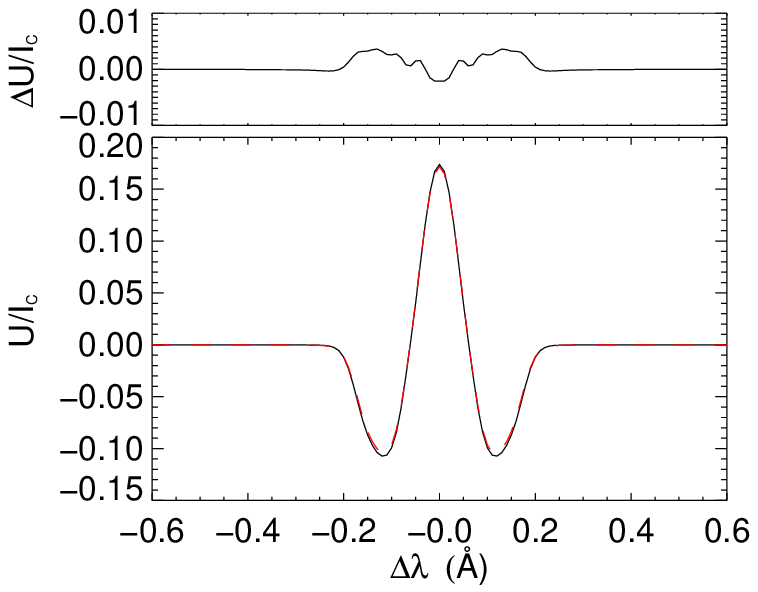}
\includegraphics[width=4.3cm]{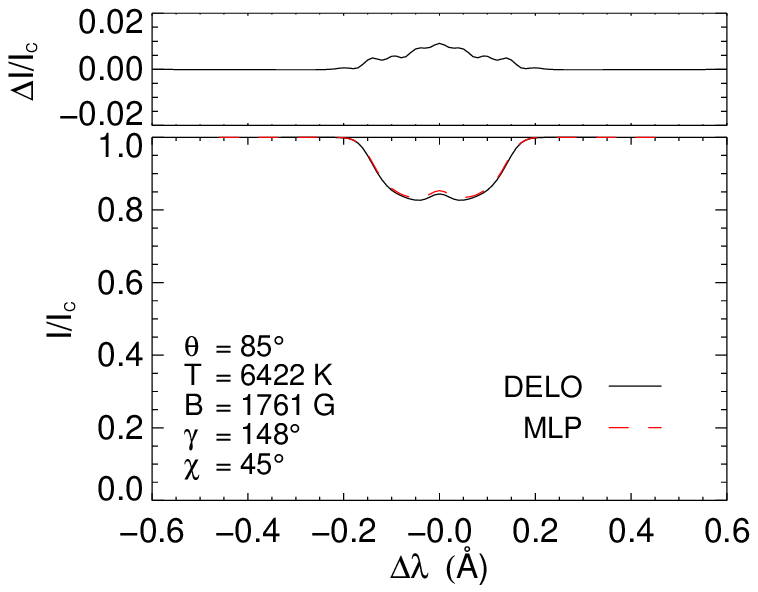}
\includegraphics[width=4.3cm]{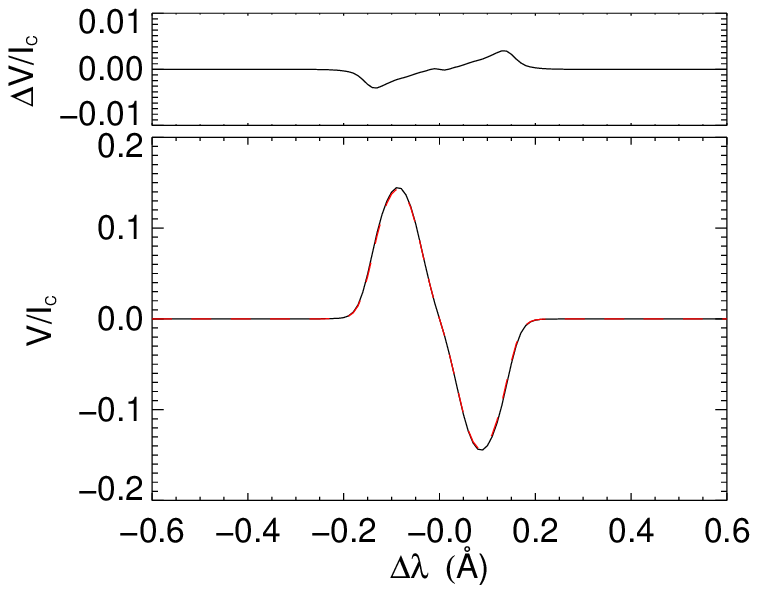}
\includegraphics[width=4.3cm]{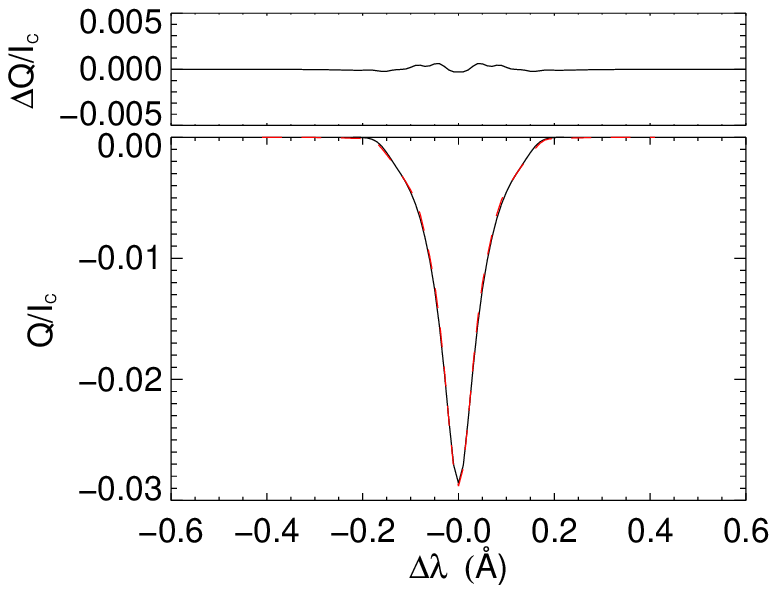}
\includegraphics[width=4.3cm]{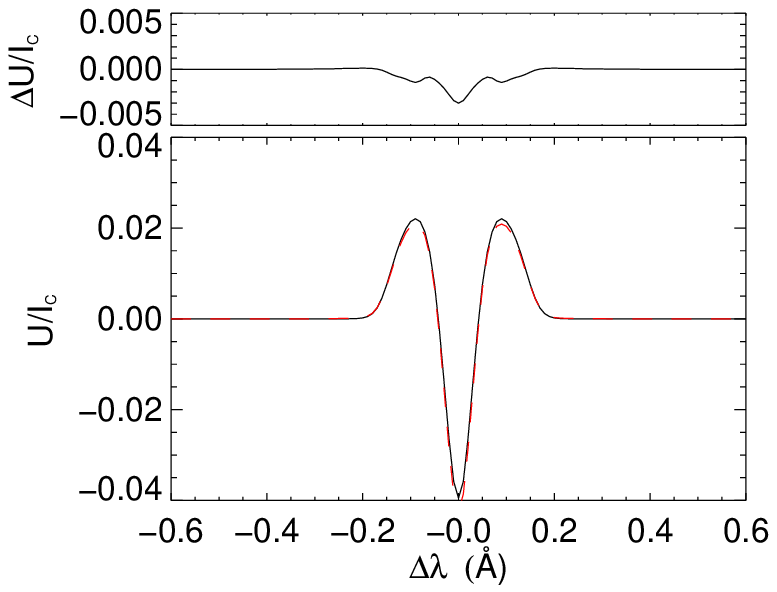}
\caption{Each row shows the set of the four Stokes profiles ($I$,$V$,$Q$,$U$),
synthesized for a particular realization of the free parameters. The particular values 
of the free parameters are always specified in the Stokes $I$ plot (first column).
Profiles calculated with the full numerical solution (denoted by DELO), are drawn with solid black lines,
while the profiles calculated with our MLP method shown in dashed red lines. Difference plots
for each Stokes parameter are provided again on top of each profile plot.}
\label{Fig:8}
\end{figure*}

To obtain yet another impression about the accuracy, we have compiled a  short list of representative Stokes 
$I$ , $V$, $Q$, and $U$ profiles in Fig. \ref{Fig:8} calculated under various combinations for the 
LOS angle of the atmosphere $\theta$,
the effective temperature of the model atmosphere $T_{\rm{eff}}$, magnetic field strength $|B|$, the magnetic field inclination
$\gamma$ and the magnetic field azimuth $\chi$. Each row in Fig. \ref{Fig:8} shows the set of four Stokes profiles calculated under
a specific atmospheric conditions. The atmospheric parameters are specified for each row in the Stokes $I$ plot (first column)
The profiles obtained by the full numerical solution (denoted by DELO) are
shown in solid black lines, while the MLP calculations are shown in dashed red lines.
For all conditions (set of free parameters) the plots show a very good agreement between 
the full numerical solution and the MLP calculation.

\subsection{Disk integrated Stokes profiles}

\begin{figure}[t!]
\centering
\includegraphics[width=8.3cm]{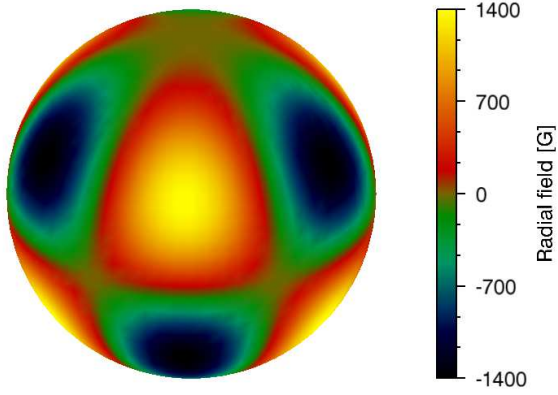}
\caption{The radial component of the surface field for the spherical harmonic test star, 
with $l$=4, $m$=3. The effective temperature of the
atmospheric model is 5000 K and the peak field strength is $\pm$ 1400 G.}
\label{Fig:9}
\end{figure}

To evaluate the accumulated effect of a disk-integrated spectrum we 
have also calculated 
the resulting profiles for two synthetic test star models within the \emph{iMap} code.
The first model consists of a potential field where the 
individual contributions to the magnetic field vector are described 
by spherical harmonics \citep{Alt69}.
For the Legendre parameter, we have chosen $l$=4 and $m$=3, which account for
a sufficient complex surface distribution of the field strength, inclination, and azimuth. 
For the magnetic field strength, we have assumed a peak value of 1400 G.
The surface distribution for the radial field component is shown in Fig. \ref{Fig:9}.

For the synthesis we used a rotational velocity with a $v sin i$ of 26 km/s. Furthermore,
we again used the iron line \ion{Fe}{i} $\lambda$ 6173 \AA\ and a Kurucz model atmosphere 
(log g = 4.0) with an effective temperature of 5000 K. 
The static parameters are those listed in Table \ref{Table:1}. The synthetic spectra are calculated 
with a surface resolution of 3 $\times$ 3 degree, which results, for the visible hemisphere, in 3600 surface
elements. For each of these surface elements the underlying model atmosphere is
adjusted with respect to the local LOS angle $\theta$ to correctly model the 
center-to-limb variation (limb darkening).

\begin{figure}[t]
\centering
\includegraphics[width=4.45cm]{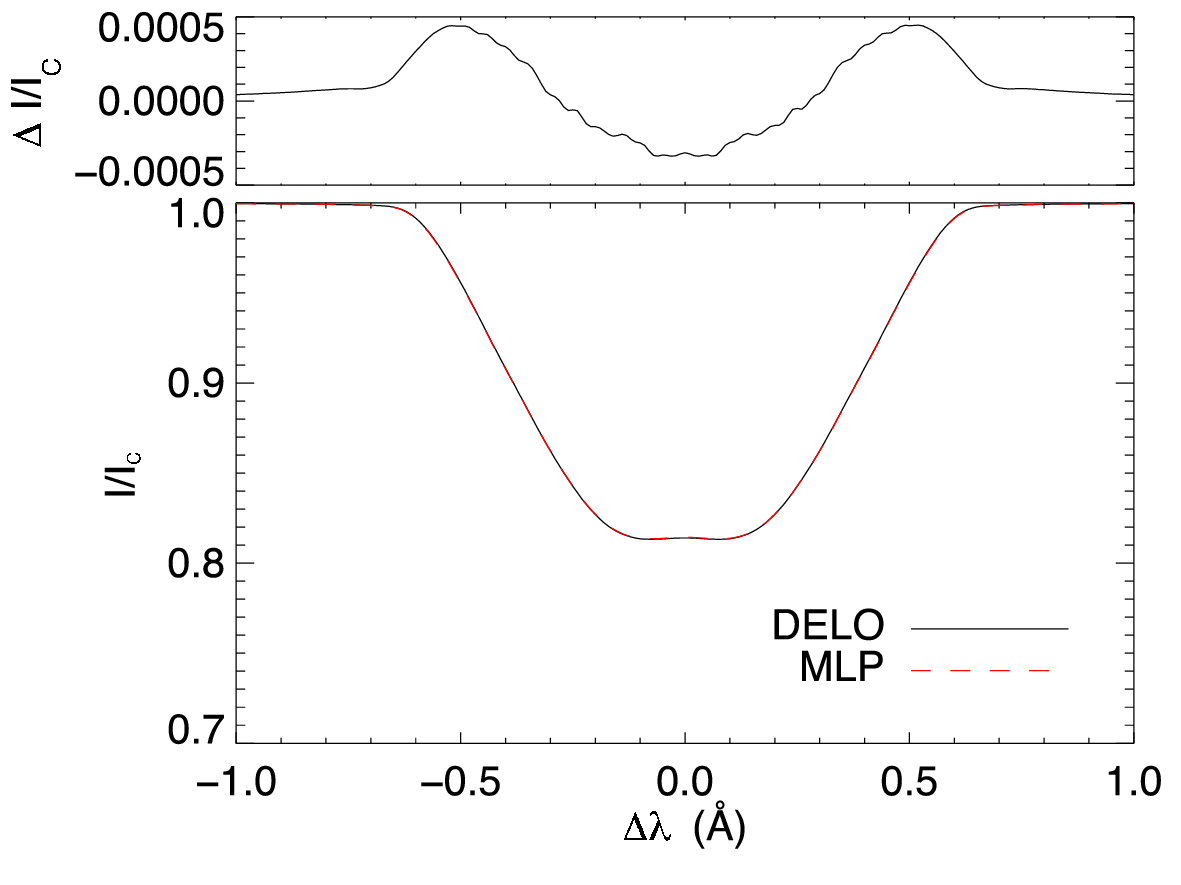}
\includegraphics[width=4.45cm]{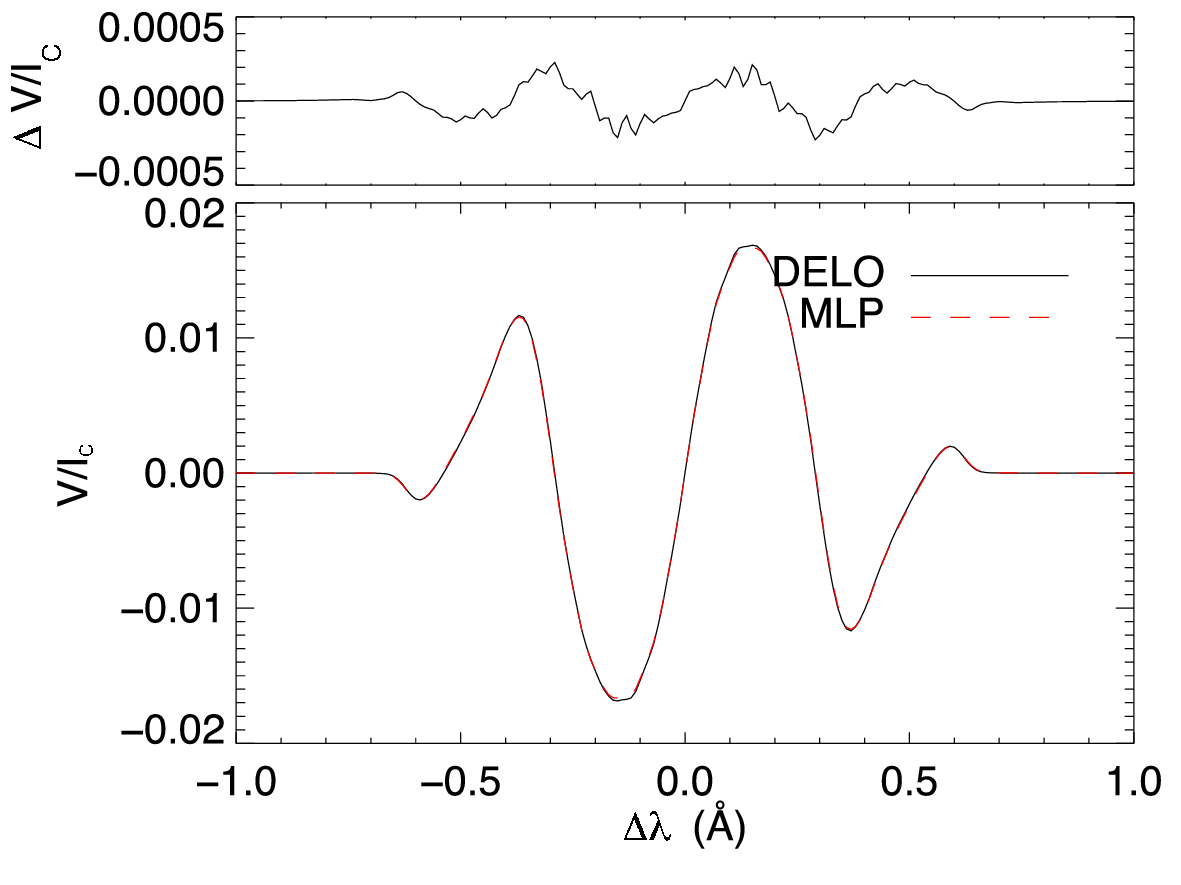}
\includegraphics[width=4.45cm]{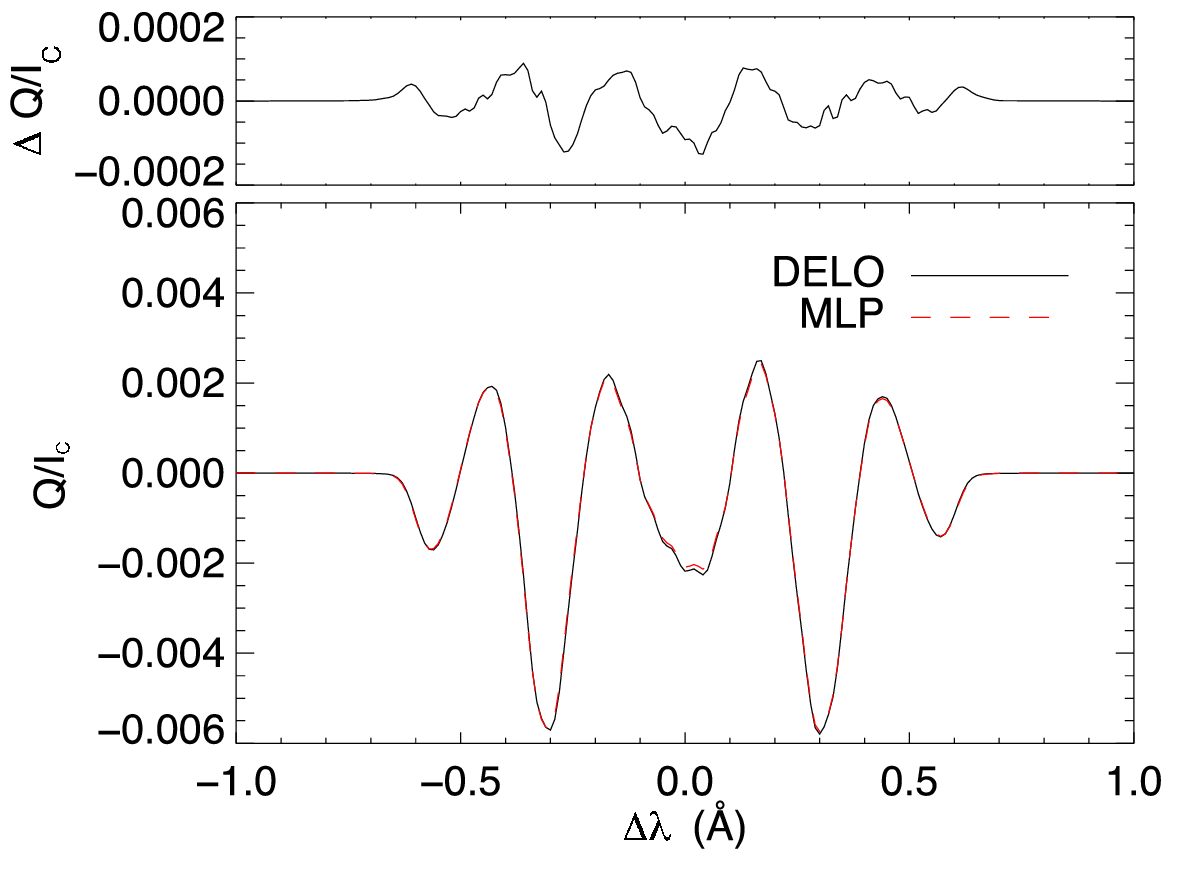}
\includegraphics[width=4.45cm]{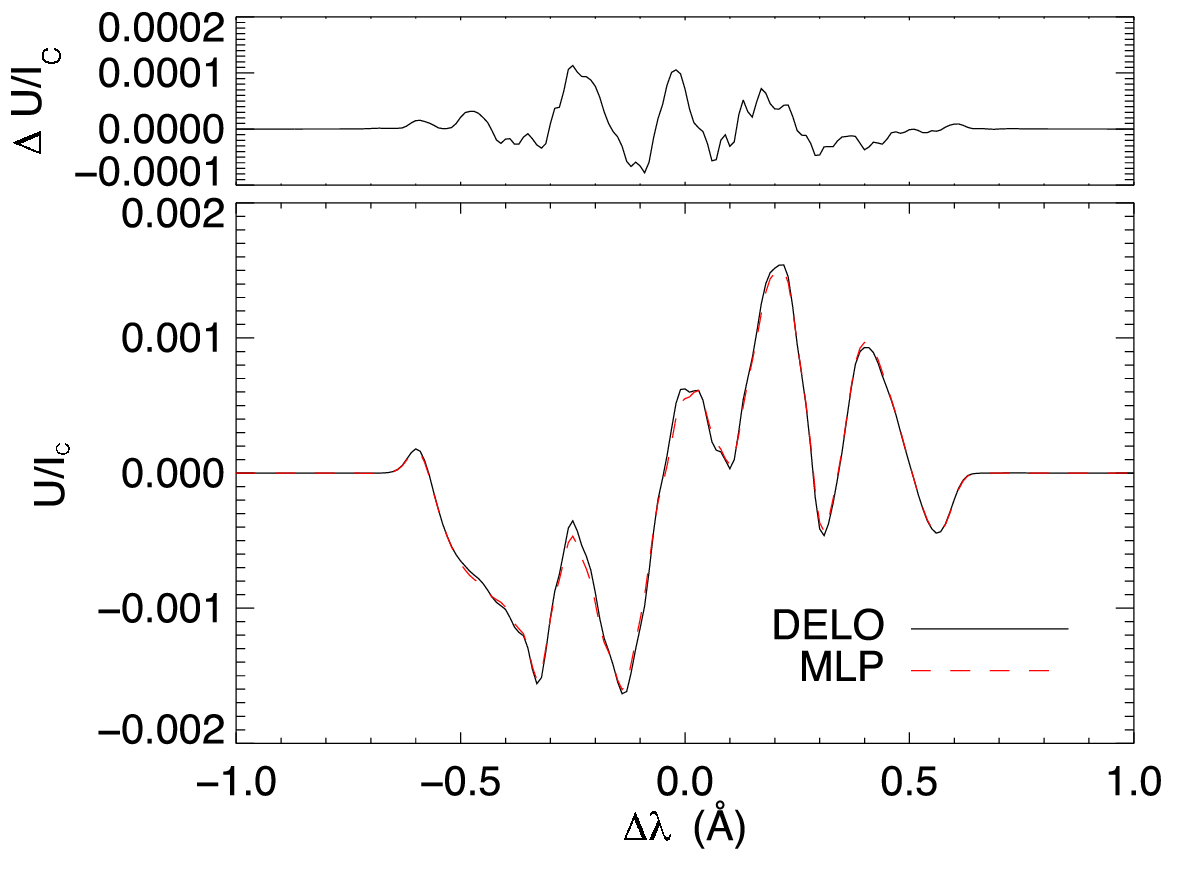}
\caption{The disk-integrated Stokes spectra for the spherical harmonic test star. Also, for the 
disk-integrated Stokes spectra, we see very good agreement between the numerical solution
(DELO, solid black line) and the MLP synthesis (dashed red line)}
\label{Fig:10}
\end{figure}

On the basis of the stellar model, the disk-integrated spectrum is calculated within
\emph{iMap}, one run with the full numerical solution and another with the MLP synthesis module. 
In Fig. \ref{Fig:10} the disk-integrated Stokes spectra are shown for the spherical harmonic model.
These plots show very good agreement between the numerical solution denoted by DELO and drawn with 
solid black lines, and the MLP synthesis shown in dashed red lines.
No accumulating errors are visible, and the errors are in fact equal or even smaller than the mean errors derived
from the statistical evaluation in Sect.\ref{Local}.
The rms error is 0.09 \% for the Stokes $I$ profile, 0.18 \% for the
Stokes $V$ profile, 0.36 \% for the Stokes $Q$ and 0.72 \% for the Stokes $U$ profile.
\label{Sect:5}
\begin{figure}[t!]
\centering
\includegraphics[width=8.3cm]{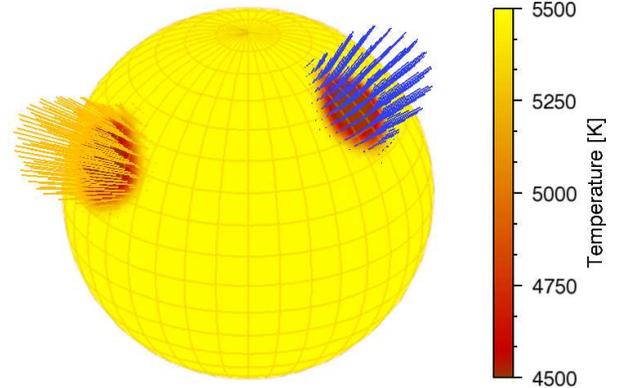}
\caption{A two-spot test star with a bipolar spot configuration of $\pm$ 1500 G. The 
effective temperature for the quiet surface is 5500 K and 4500 K for the spots.}
\label{Fig:11}
\end{figure}
\begin{figure}[t]
\centering
\includegraphics[width=4.3cm]{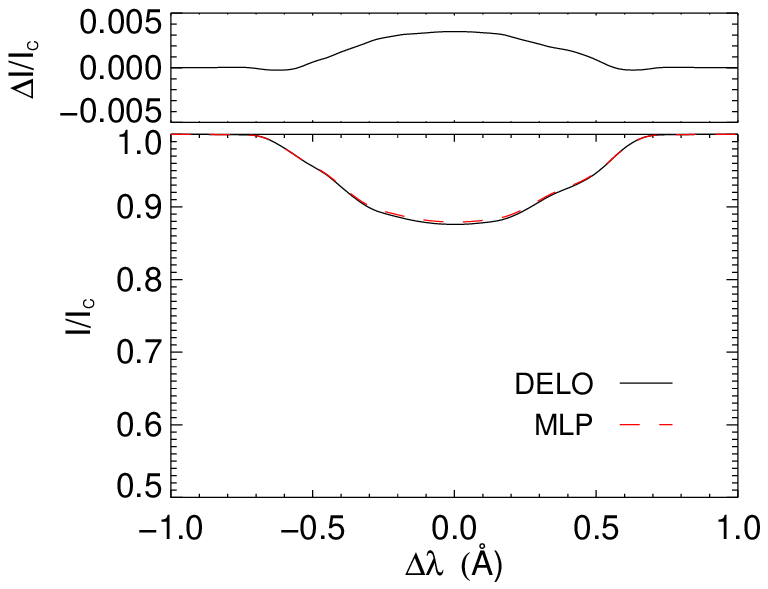}
\includegraphics[width=4.3cm]{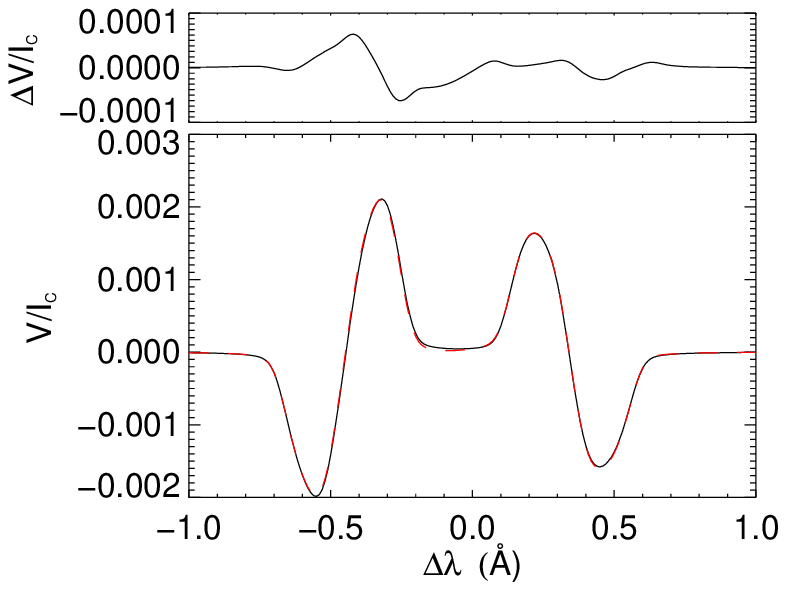}
\includegraphics[width=4.3cm]{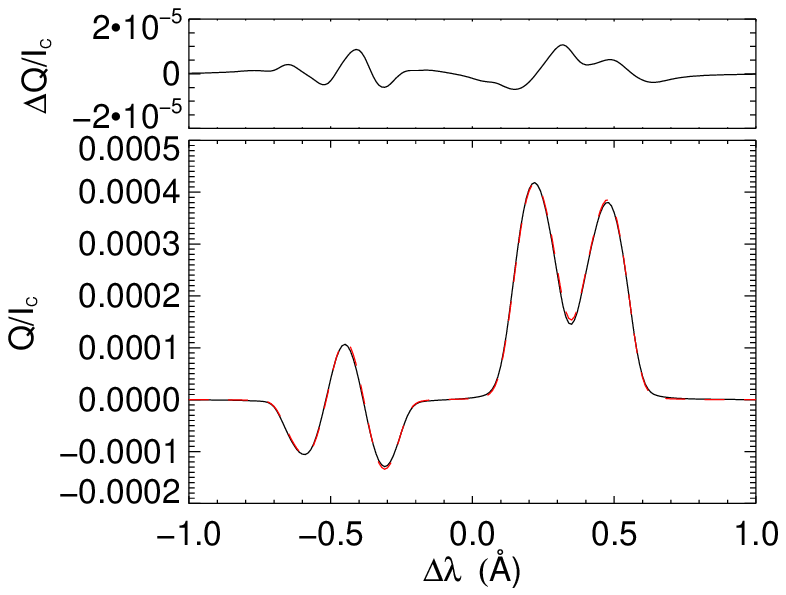}
\includegraphics[width=4.3cm]{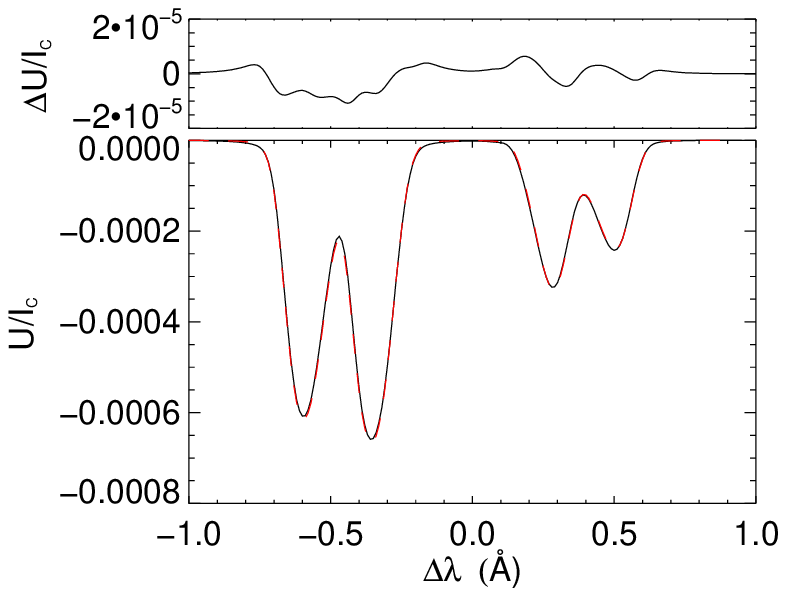}
\caption{The disk-integrated Stokes spectra for the two spot test star. Again, an impressive
agreement between the correct numerical solution (DELO, solid black line) and the MLP synthesis (dashed red line)
is achieved.}
\label{Fig:12}
\end{figure}   
To provide another model and to test the joint effect of 
a temperature and a magnetic field variation, we construct with \emph{iMap} a two spot model star.
For the quite stellar surface, we used a Kurucz model atmosphere (log g = 4.0) 
with an effective temperature of 5500 K and for the two spots a model that is 1000 K cooler (4500 K). 
The maximum field strength of the spots is set to 1500 G and the field orientation is such that
the two spots have a radial field of different polarities.
The $vsin i$ is 35 km/s and the disk integration is performed 
with a surface resolution of 5 $\times$ 5 degree.
To provide a smooth continuous gradient in the temperature as well as in the magnetic field 
within the two spots we have also smeared the field and temperature distribution on the surface 
with a 2-dimensional Gaussian function. 
Note that in this model the relative weighting of the local profiles is also affected by the 
temperature variation and the corresponding continuum intensities, which therefore requires
an accurate calculation of the individual continuum intensities by the MLP.
The model star is shown in  Fig. \ref{Fig:11} where we have color coded the surface temperature
and depicted the field lines within the spots by blue lines for positive and yellow lines 
for a negative polarity. 

The resultant Stokes profiles for the disk integration are shown in
Fig. \ref{Fig:12}, where we can see again an exceptional agreement between the numerical and the
neural network calculation.  The rms errors are 0.12 \% for the Stokes $I$ profile, 0.37 \% for the
Stokes $V$ profile, 0.44 \% for the Stokes $Q$, and 0.43 \% for the Stokes $U$ profile.

 %__________________________________________________________________

\section{Improvement in speed}

One of the main motivations for our proposed approach 
was of course to find a suitable approximate method which facilitates 
a fast synthesis of Stokes profiles. 
Once our MLPs are trained and the network weights are frozen, 
the network structure represents a simple form of a composite mathematical function (see Sect. \ref{Sect:3}). 
The MLP networks can then readily be implemented in a quickly-evaluated programming language 
or stored on an even faster to evaluate hardware chip (e.g., EPROM). 
Both approaches (whether software or hardware) have significant potentials to allow for an efficient parallelization.
For our purpose we implemented the five trained MLP network structures 
in the programming language C++ on a single processor desktop computer.

To evaluate the increase in speed provided by our ANN approach,
we made the following benchmark test.
Both methods, the numerical integration of the polarized radiative transfer equation
with DELO (which is one of the fastest numerical methods) and the MLP synthesis with the 
MLP networks, are used to calculate 10,000 times the full set of Stokes profiles 
($I$,$Q$,$U$,$V$) as well as the local continuum intensity. The profiles are calculated in a 
wavelength range of $\pm$2 \AA\ around the line center with a
spectral resolution of 0.01 \AA\ , which gives 401 wavelength points in total for each 
Stokes parameter and each run. The spectral line is again the iron line \ion{Fe}{i} $\lambda$ 6173 \AA\ .
The DELO routine integrates the spectral lines within a depth range of 
log($\tau_{5000}$) = 1.0 and log($\tau_{5000}$) = -5.0.
The input parameters for the numerical integration and the MLP method are
chosen randomly, but prior to the calculation and were kept fixed for the entire
benchmark cycle. By this we allow the numerical integration method 
to calculate and compile the absorption matrix Eq. (\ref{Eq:4}) and (\ref{Eq:5}), once and 
for all in advance. Moreover, it avoids the time consuming reading
of model atmospheres and line parameters.

The benchmark was performed on several personal computers with varying performance characteristics
(processors ranging from 1.7 GHz to 3.2 GHz and RAM from 512 MB to 1 GB).
The test runs for the ANN/MLP method takes between 2.1 and 4.2 seconds while the 
conventional numerical calculation of the polarized radiative transfer with the DELO method 
takes between 2820 to 3610 seconds.
This gives an improvement in speed by factors of 860 to 1340 !
It is this significant improvement in speed -- by three orders of magnitude --
which facilitates a complete disk-integration of a high-resolution stellar model within a few seconds.
With this accelerated Stokes profiles synthesis, 
the complete cycle of forward and inverse calculation for ZDI applications
now become possible within a reasonable time frame. 

%__________________________________________________________________

\section{Discussion and conclusions}
\label{Sect:6}
In this paper we have introduced, described and evaluated a novel
LTE Stokes spectrum synthesis method based on ANNs.
Artificial neural networks are used to approximate the LTE polarized
radiative transfer by modeling the functional relationship between the 
most important atmospheric parameters and the corresponding local Stokes spectra.
The ANN approach with MLPs is based on an extensive training database of calculated
synthetic Stokes profiles.
Because of this crucial dependence for the network training we first
assessed the accuracy of our polarized radiative transfer module
integrated in our ZDI code \emph{iMap} by a detailed comparison with the 
existing synthesis code C{\small OSSAM} in Sect. \ref{Sect:3}.
Both codes exhibit congruent results for several composite values like
the continuous absorption coefficient and broadening parameters.
Moreover, both codes show a very good agreement in all calculated 
Stokes profiles. 
By verifying the agreement between \emph{iMap} and C{\small OSSAM},
we also indirectly established an inter-agreement between \emph{iMap} and 
I{\small NVERSE}10 as well as Z{\small EEMAN}2, as derived from the 
inter-agreement study of \citet{Wade01}.

In Sect. \ref{Sect:4}, we introduced the concept of our proposed method and described
the data preparation with a PCA method as well as the training process for the ANNs.
A detailed evaluation of our ANN synthesis revealed the remarkable accuracy with which
the ANN approach is able to approximate the polarized radiative transfer process. 
Stokes profiles of the spectral line \ion{Fe}{i} $\lambda$ 6173 \AA\ are calculated with 
high accuracy for different configurations of the atmospheric input parameters (effective temperature, 
magnetic field strength, field inclination, field azimuth, and LOS angle).
The ANNs have clearly been able to \emph{learn} the complex 
non-linear mapping between the atmospheric input parameters and the resulting Stokes profiles.
Although we have restricted in this work the number of variable input parameters to 5, there is
no principle limitation to extend the number of atmospheric input parameters. 
Moreover, the limits that we have set for the parameter ranges (see Table \ref{Table:1}) are also
not a principle restriction of the ANN approach but rather driven by the constraints
to cope with the training and evaluation database in this study.

The ANN approach to spectral line synthesis is also not limited to
one spectral line, the ANN method can easily be extended to a wider wavelength range
to cover several neighboring spectral lines or successive regions of interest in the wavelength 
domain.
Another interesting possibility, offered by the ANN approach is the 
modeling of co-added line profiles as used for polarized spectral line extraction and 
reconstruction algorithms like LSD \citep{Donati97c} or by the 
PCA method  \citep{Carroll07,Marian08}. 
Instead of synthesizing a huge number of individual Stokes profiles to model, the
co-added (and therefore noise reduced) mean line profile, an ANN synthesis can be used to 
directly calculate this co-added line profile. This would offer a vast improvement in 
speed and accuracy.

It is also important to note that since the ANN approach is based on the accurate 
numerical modeling of line profiles, blends can be fully accounted for with this approach.
Even heavily blended line profile can be accurately synthesized as long as the 
training database properly accounts for these blends. 

One of the main reasons to approximate or even to entirely bypass 
a full radiative transfer approach in DI or ZDI applications,
is the enormous requirement for massively calculating local Stokes profiles 
during the inversion process. 
In our benchmark tests, we could show that our ANN approach is three orders of magnitudes 
faster then the conventional numerical integration scheme.
This enormous acceleration and the impressive accuracy makes the ANN synthesis approach
a viable and promising alternative for polarized radiative transfer calculations
and facilitates a complete radiative transfer driven approach to DI and ZDI as well as
for large scale Stokes profile inversions. 

%__________________________________________________________________

\begin{acknowledgements}
The authors would like to thank the referee A. Asensio Ramos for his constructive and helpful
comments, which helped to improve this manuscript.  
\end{acknowledgements}

%__________________________________________________________________

\end{document}